\documentclass{article}

\usepackage{arxiv}
\usepackage{centernot}
\usepackage[utf8]{inputenc} 
\usepackage[T1]{fontenc}    
\usepackage{url}            
\usepackage{booktabs}       
\usepackage{amsfonts}       
\usepackage{nicefrac}       
\usepackage{microtype}      
\usepackage{lipsum}
\usepackage{graphicx}
\usepackage{verbatim}
\usepackage{latexsym}
\usepackage{subcaption}
\usepackage{setspace}
\usepackage{float}
\usepackage{amsmath}
\usepackage{braket}
\usepackage{mathtools}
\usepackage{tensor}
\usepackage[]{algorithm2e}

\usepackage{wrapfig,lipsum}
\usepackage{multicol}

\title{Evaluating structural edge importance in temporal networks}

\usepackage{authblk}
\author[1,2]{Isobel Seabrook}
\author[1]{Paolo Barucca}
\author[1, 3, 4]{Fabio Caccioli}
\affil[1]{Department of Computer Science, University College London, London, United Kingdom}
\affil[2]{Financial Conduct Authority, London, United Kingdom}
\affil[3]{Systemic Risk Centre, London School of Economics, London, United Kingdom}
\affil[4]{London Mathematical Laboratory, London, UK}
\date{}                     
\begin{document}
\maketitle

\begin{abstract}

To monitor risk in temporal financial networks, we need to understand how individual behaviours affect the global evolution of networks. Here we define a structural importance metric -- which we denote as $l_e$ -- for the edges of a network. The metric is based on perturbing the adjacency matrix and observing the resultant change in its largest eigenvalues. We then propose a model of network evolution where this metric controls the probabilities of subsequent edge changes. We show using synthetic data how the parameters of the model are related to the capability of predicting whether an edge will change from its value of $l_e$.
We then estimate the model parameters associated with five real financial and social networks, and we study their predictability.
These methods have application in financial regulation whereby it is important to understand how individual changes to financial networks will impact their global behaviour. It also provides fundamental insights into spectral predictability in networks, and it demonstrates how spectral perturbations can be a useful tool in understanding the interplay between micro and macro features of networks.  
\end{abstract}

\section{Introduction}
Understanding how individual edges in a network influence its structure and evolution is important in a range of applications. Considering financial networks, network structure has implications for financial stability \cite{Nier}, market efficiency \cite{Silva} and consumer safety \cite{Allen}. Identification of players to monitor more closely is of paramount importance to regulators and policy makers, with many attributing the severity of the 2008 crisis to systemic flaws in the banking ecosystem \cite{Haldane}. Our research focuses on understanding how individual edges affect the structure of networks, and how this relates to network stability and evolution. We present a brief review of related literature, first considering individual effects on network structure, then those that link network structure to stability and systemic risk, before considering how network structure relates to temporal evolution. We then define a measure for structural edge importance $l_e$, and we propose a model for network evolution in which an edge's importance can be indicative of future changes. Our results show that $l_e$ values are higher for edges which appear to play a more important structural role, and that subsequent changes occurring in the real networks analysed depend to some extent on the value of $l_e$. 
\subsection{Individual effects on network structure}
The effect an individual node or edge can have on a network's structure depends not only on the scale of its activity, but also on its position within the network, and the activity of neighbouring nodes and edges. Understanding these interrelations remains one of the key challenges in network science.

Recently, structural node importance has gained a large amount of attention due to its relevance in use cases across a wide range of fields \cite{Ren}. Methods have predominantly focused on network spectra, in order to illicit structural information from the network adjacency matrix. This includes numerous studies of epidemic processes, in which it is intuitive that the removal of a node that acts as a bridge between communities can be used to stem the spread of a disease, leading to significant effort being taken to understand the influences of community structure on epidemic spreading \cite{Ren,Stegehuis,Ren1}. Similar applications include preventing network-based attacks \cite{Lai,Xuhui} and understanding and actioning on the spread of gossip in society \cite{Moreno}. This idea of network resilience is often approached from the angle of percolation theory, in which the percolation threshold governing the appearance of a giant component is often related to the leading eigenvalue of the adjacency matrix \cite{Restrepo, Bollobas}.  An alternative lens is taken by Wang et. al. \cite{Wang} who make use of the observation that the spectrum of the adjacency matrix gives an indication of community structure. In noting that for a network with c strong communities, the c largest eigenvalues of the adjacency matrix are significantly larger than the others, they follow a perturbation based approach to define node importance as the relative change in the c largest eigenvalues of the upon the node's removal. 
Similar to Wang et. al., Lü et. al. \cite{Stanley} propose a universal structural consistency index for a network-based on perturbing the adjacency matrix and demonstrate that this index is a good index for link predictability. Our work considers the same central concept of applying perturbations to the adjacency matrix and focusing on the change in the leading eigenvalue, however differs in that we are working at a lower level of granularity by proposing an edge-based measure, rather than node or network-based. 

Many works in the financial literature focus on node specific influence on stability. For example, Battiston et. al. \cite{Battiston} define a node ranking coined DebtRank, which takes recursively into account the impact of distress of an initial node across the whole network. Their measure amounts to the fraction of the total economic value in the network that is potentially affected by the distress or default of a specific node. They applied their method to a network of loans from the Federal Reserve to financial institutions between 2008 and 2010, enriched with equity investment relations, and found a strongly connected core of 22 institutions which all became too systemically important to fail at the 2008 crisis peak. They demonstrated the effectiveness of their node ranking in comparison to other centrality measures, and found that it was the only measure to deliver a clear response well before the crisis peak. However, their method specifically considers the case of distress propagation, and does not explicitly measure how an individual node or edge affects the structure of the network in general. Barucca et. al. \cite{Barucca} investigate whether a change to few selected banks in the network of the e-MID\footnote{e-MID is the Italian electronic market for interbank deposits, a platform for trading unsecured money-market deposits.} market can affect the large scale structure of a network through node removal or degree mutation, and comparing the network structure that results to the original. 

Although the bulk of the attention has focused on importance of actors in networks, Helander et. al. \cite{Helander} propose a method for characterising the relative importance of an edge, which they refer to as edge gravity. Edge gravity measures how often an edge occurs in any possible network path. They show that important edges are not necessarily adjacent to nodes of importance as identified by standard centrality metrics, and they also observe that high centrality nodes often have their centrality over-represented by being adjacent to `edges to nowhere'. Similar path-based methods include the $BCC_{MOD}$ (Betweenness Centrality and Clique Model) proposed by \cite{Yu}, which weights the importance of the two nodes forming the endpoints of the edge with the number of cliques containing the edge. Their method outperforms several well-known methods including Jaccard
coefficient and betweenness centrality in identifying critical edges both in network connectivity and spreading dynamic. In our work we define the importance of an edge in terms of the change that a small perturbation on the edge would induce in the leading eigenvalue of the weighted  adjacency matrix of the network. While other definitions could be considered, we focus here on the leading eigenvalue because it determines for instance the stability of spreading processes on social networks \cite{Wang_eigs, Pei}, or financial shocks on inter-bank networks \cite{Bardoscia}.
 Our methods contrast the above-mentioned path-based approaches by instead considering a network spectrum-based approach, however both approaches show strong connections to node centrality measures; as shown in section \ref{sec:method}, an approximation to the network eigenvalue derivative is proportional to the product of the constituent nodes' centralities. In addition, our research focuses on the temporal behaviour of the network in relation to structural importance, for which future work could consider using alternative measures of structural importance to understand the expected temporal behaviour.

\subsection{Network structure in relation to stability and systemic risk}
Increasing complexity and stability are inextricably linked, with works as early as May's investigations into ecosystems with increasing biodiversity highlighting the relationship \cite{May}. 
In the context of financial markets, although market integration and diversification are widely believed to play a stabilising role \cite{Restoy, Samuelson}, Bardoscia et. al. \cite{Bardoscia1} demonstrated that two factors of increasing complexity, namely increasing the number of institutions (nodes) and contracts (edges) in an interbank network can drive the system to instability. 
Similarly, Markose et. al. \cite{Markose} present the idea of institutions being `too interconnected to fail' through an exploration of the structure of the US CDS market. They consider an empirical network constructed from market shares, and make use of the May-Wigner condition for stability \footnote{The May-Wigner condition for stability is a critical threshold below which any random network has a high probability of stability, and is defined as $D<\frac{1}{Ns^2}$ where $D$ is the network diameter, N is the number of nodes and $s$ is the strength  of  average  interactions  between  nodes} in comparison to a random network. They show that although the CDS structure shows better outcomes than a random network when subject to shocks, the demise of any one big player will bring down other big players. Caccioli et. al. \cite{Caccioli1} 
showed in a theoretical exploration that uncontrolled proliferation of financial instruments can lead to large instability in markets, and suggest potential interventions such as the introduction of a Tobin tax \cite{Tobin}, which is shown by Bianconi et. al. to have a stabilising effect \cite{Bianconi}. Related to this, Brock et. al. \cite{Brock} used `arrow securities' as a proxy for more complicated hedging instruments, and found that these incentivise construction of larger positions, resulting in a reinforcement effect due to large gains/losses as a result of being on the `right' or `wrong' side of the market. They showed that this is associated with greater instability, and also that the primary bifurcation parameter, marking the onset of instability, occurs earlier when there are more arrow securities. In contrast to the majority of the data centric financial literature which focuses on interbank trading, Bardoscia et. al. \cite{Bardoscia} analysed UK Trade Repository data, which includes all transactions occurring through a Central Counterparty clearing house (CCP) in the UK. Considering a snapshot of the open positions on a single day for interest rate derivatives, FX derivatives and credit default swaps as a three layered network, they compared a ranking derived from the centrality measures to a ranking derived from modelling the network's response to liquidity contagion, looking at how shocks propagate across the network and translate into payment deficiencies across the different markets. The model considers the stress faced by an institution - the difference between all payments it is required to make and all payment inflows from counterparties, and allows stress to spill over between the layers. They found that centrality measures can be used as a proxy for the vulnerability of financial institutions.

\subsection{Network structure in relation to temporal evolution}

To understand how networks evolve across time, many researchers have focused on studying the mechanisms for network growth, and defining network models to understand the origin of observed properties of real networks \cite{Aste1, Mazzarisi,Barucca1, Barucca3,Sora}. These include the Barabasi-Albert model \cite{Barabasi}, which demonstrates that scale-free degree distributions observed in real networks can be explained by the presence of growth and preferential attachment in the network evolution. Falkenberg et. al. \cite{Falkenberg} present a simple adaptation to the Barabasi-Albert model, in which new nodes attach to nodes in the existing network in proportion to the number of nodes one or two steps from the target node. This results in an implicit time dependence, which arises from a node's attractiveness being dependent on its local environment which changes as the network evolves. Central to their model is the idea that network structure and temporal evolution are inherently linked, however their model is limited to the influence of local environment.  Others focus on considering temporal networks as multilayer networks, in which one can account for the fact that connectivity patterns in different layers can depend on each other. Bazzi et al. \cite{Bazzi} proposed a generative model which explicitly incorporates a user-specified dependency between layers that is flexible enough to incorporate complex interlayer relationships such as dependencies between a layer and all layers that follows, incorporating memory effects into the model. 
A handful of studies have attempted to link global network structure to temporal evolution, such as Peixoto et. al. \cite{Peixoto}, who suggest dynamical variation of the degree-corrected stochastic block model that is capable of finding meaningful large-scale temporal structures in real-world systems and predict their temporal evolution. Their method works with both discrete and continuous time representations, making it versatile to a range of applications. Watts et. al. \cite{Watts} consider semi-random `small world' networks and show that the dynamics are an explicit function of the network structure, and also show find an enhanced propagation speed for small world networks. 

A common and general framework for network growth is the fitness model, in which each node has associated with it a time independent `fitness' which represents its propensity to attract links, as proposed by Barabasí and Bianconi \cite{Bianconi1}. They find that different fitnesses results in multiscaling in the dynamic evolution, or in other words that the time dependence of a node's connectivity depends on the fitness. Attempts have been made build on this model in order to understand the origins of network dynamics, such as a recent study by Kobayashi et. al. \cite{Kobayashi}. They find that population and activity dynamics are sufficient to explain two types of scaling empirically observed in real networks, however their methods do not explicitly allow for different roles to be captured within a network, by assuming a uniform distribution of fitness parameters. In our research, we explore instead how an edge level quantity derived from the spectrum of a network can similarly be used to determine which edges change in the network. We present methods for estimation of parameters which control both the overall activity in the network, as well as the bias to change for edges with a larger structural importance, and we show how these reproduce behaviours observed empirically.

In the following sections, we look to address two questions: Can we quantify the extent to which an edge affects the overall network structure, and does this provide information on the network's temporal evolution? We know from the above that network structural information can be gained from the network spectra, both from the observation that the threshold for the appearance of a giant component in a network relates to the leading eigenvalue, and in that the number of communities can be determined from the number of well separated eigenvalues. We also see that the leading eigenvalue provides an indication of stability in terms of dynamical processes occurring on the network. Our aims are to understand node importance in terms of network structure and stability, so we thus look to capture both of these in our analysis through considering the derivatives of the network's leading eigenvalue with respect to individual nodes or edges. We present evidence that this measure could be a useful indicator in understanding temporal changes in network structure, and we present the results of its application to five real networks. Our main results demonstrate that the elementwise derivative of the leading eigenvalue ($l_e$) can be predictive of subsequent change for five different networks analysed, and that predictability can be related to the specific realisation of two parameters, $\alpha$ and $\rho$ in the network evolution model in which edges change with probability $\alpha l_e^{\rho}$. This has potential implications for stability, as a system experiencing more changes to edges of structural dominance could see a reinforcing effect, leading to an unstable system. These methods could be useful in classifying financial asset systems to inform regulation activities and policy making. We further show that the scale of resultant changes can be related to the realisation of two additional parameters $\beta$ and $\gamma$, again with potential stability implications.
\section{Methodology}
\label{sec:method}
\subsection{Definition of temporal networks}
Traditionally, network analytics has focused on static representations of networks, either looking at single snapshots in time, or considering a projection of the time dimension onto a static view by aggregating the links in a time window. In doing so, some, or all, of the temporal information about the network is lost. 

However, recently, there have been developments in the modelling of systems as temporal networks, for which the system is represented by a contact sequence $(i,j,t)$, where $i$ and $j$ constitute the vertex set $V$ at time $t$. This representation also allows for edges that take time to traverse, or contracts completed after a duration $\delta t$ by representing the contact sequence as $(i,j,t, \delta t)$ \cite{Holme}.
Since we are considering transactions as instantaneous, we are not interested in transmission time for edges, and we are considering applications where time is discretised, we can formally define a temporal graph $G^w_t (t_{min}, t_{max})$ as in \cite{Tang} as the ordered sequence of graphs
\begin{equation}
    G_{t_{min}} , G_{t_{min}+w},
. . ., G_{t_{max}}
\label{temp_graph}
\end{equation} 
where $w$ is the size of the time aggregation (e.g. daily). Element $A^s_{ij}$ of the adjacency matrix at time $s$ is 1 if and only if there
exists a link between $i$ and $j$ in $G_t$, $t \leq s \leq t + w$. 

This differs from a time sequence of static graphs in that the edges in the temporal network need not be transitive, i.e. 
$A \rightarrow B , B \rightarrow C \centernot\implies  A \rightarrow C$, 
and it also allows for time to be continuous, meaning the full topological structure and correlations are captured. Temporal networks can be extended to include weights associated with the edges.

\subsection{Central concept - eigenvalue derivatives as a measure of importance}
\label{sec:eig_derivs}
For a given graph $G_t(V,E)$ with adjacency matrix $A_{ij}^t$, the eigenspectrum of $A_{ij}^t$ is the set of eigenvalues $\lambda$ that satisfy the equation 
\begin{equation}
    \mathbf{A x} = \lambda \mathbf{x}
\end{equation}

By observing changes in the eigenspectrum of a graph, we can gain an insight into structural changes. As we are looking at network snapshots across time, we have a `time series' of graphs and we can consider the change in the leading eigenvalue between successive time snapshots,
\begin{equation}
    \Delta \lambda = \lambda(A^{(t+1)})-\lambda(A^{(t)}) \approx \sum_{ij} \frac{\partial \lambda}{\partial A_{ij}}\Delta A_{ij}
    \label{initial theory}
\end{equation}
where we have made a first order approximation, and the derivative is with respect to the $(i,j)$th entry of the matrix, as opposed to the entire matrix. Here $\lambda$ refers to the leading eigenvalue of the adjacency matrix. 

The two parts of equation \ref{initial theory} can be seen as a playoff between the potential of an edge to influence the structure $(\frac{\partial \lambda}{\partial A_{ij}})$ and the actual change in the network structure ($\Delta A_{ij}$). Our experiments with synthetic networks look to assess the extent to which our derivation below, which makes approximations and assumptions, captures the true behaviour. The first term measures the sensitivity of the eigenvalue to changes in an individual edge, which we refer to as the structural importance of an edge and denote by $l_e$. We derive approximations for $l_e$ in equation \ref{undirected intermediate} for the undirected case by taking a perturbation theory approach. Although not explicitly explored in this paper, we also present equation \ref{directed intermediate} for the directed case: 
\begin{equation}
   l_e = \frac{\partial\lambda}{\partial A_{ij}}=2\lambda_{0,i}\lambda_{0,j}
    \label{undirected intermediate}
\end{equation}
\begin{equation}
    \frac{\partial s^A}{\partial M_{ij}}=\frac{\lambda^M_{0,i} \lambda^M_{0,j}}{2s^A}
    \label{directed intermediate}
\end{equation}
where $\lambda_{0,i}$ refers to the $i$th component of the eigenvector corresponding to the leading eigenvalue, $s^A$ refers to the leading singular value of the adjacency matrix and $\lambda^M_{0,i}$ refers to the $i$th component of the eigenvector corresponding to the leading eigenvalue of $\mathbf{M} = \mathbf{A A^T}$.
Full derivations for these can be found in appendix \ref{appendix:eig_derivs}, and we validate the approximation for the undirected case in results section \ref{sec:le_validation}. We see here that both equations \ref{undirected intermediate} and \ref{directed intermediate} are proportional to the product of the eigenvector centralities of the nodes involved in the edge. Our definitions are defined in terms of the eigenvector corresponding to the largest eigenvalue, which usually has non-zero values only for the largest connected component of a network. For this reason, in this paper we restrict ourselves to exploring the giant component of the networks, however generalising these to allow for disconnected components will be considered in future work. 

We can capture the relationship between $l_e$ and subsequent edge changes by observing the distributions of $P(\Delta A=0|\ln(l_e))$ and the joint probability $P(\Delta A,l_e)$, which we explore in detail in the results sections \ref{sec: res_dyn_real} and \ref{weight_params}. Our findings from these are compared to our model for the temporal evolution of networks, which we propose in section \ref{sec:model_prop}, to assess the extent to which our model captures the true behaviour observed. 

The second term considers the changes that subsequently occur in response to the value of $l_e$. This is of significance from a stability perspective; edges that are structurally important could cause a system to become unstable by changing frequently or by a large amount. Conversely, they may also act to stabilise a system if it begins to move towards a regime of instability. This can be explored by considering temporal graphs as described by equation \ref{temp_graph}, and assuming that the evolution is Markovian. We consider this first of all in the proposal of a model for network evolution, parameterised by the extent to which $l_e$ is indicative of the propensity of an edge to change, and the scale of the resultant changes. We further assess the predictability of changes from the value of $l_e$ through the use of a logistic regression classifier, and relate the performance of this to the model parameters.

\section{Model for network evolution}
\label{sec:model_prop}
In order to understand the relation between structural importance and stability of a network over time, we need a model that captures two behaviours. The first of these is that the value of $l_e$ is indicative of the probability for an edge to change, and the second is that the size of a resultant change can be related to $l_e$.

We thus propose a model in which we can control the extent to which $l_e$ influences a subsequent edge change, both in probability of occurrence and resultant scale. Specifically, we propose a model in which the network evolution exhibits the Markovian property as in \cite{Mazzarisi,Peixoto}: 
\begin{equation}
    A_{ij}^t=V_{ij}^tA_{ij}^{t-1}U_{ij}^t+(1-V_{ij})A_{ij}^{t-1}
    \label{markov_chain}
\end{equation}
where $V_{ij}\sim \mathbb{B}(\alpha(l_e)^\rho)$ and $U_{ij}^t$ is the distribution of edge changes. Here we introduce two parameters which control the probability of an edge to change - $\rho$ which controls the level to which the value of $l_e$ influences the probability for an edge to change, and $\alpha$ scales $V_{ij}$ to ensure that it is a valid probability. A positive value for $\rho$ indicates that more important edges are more likely to change, and a negative $\rho$ would indicate the opposite. 

The simplicity of this model means that we are unable to account for edges appearing and disappearing in the network. We will look to incorporate this in future research.
\subsection{Parameter estimation in real networks}
Assuming that our data evolves according to the model in equation \ref{markov_chain}, we can use observations from real networks to estimate the most likely values of $\alpha$ and $\rho$ from the data. Following a maximum likelihood approach, we can derive estimations for these parameters, by maximising the following log-likelihood as proposed in appendix \ref{parameter_estimations}:

\begin{equation}
    \ln(L(\mathbf{x}|\mathbf{\theta}))=\sum_{e}^N k_e ln(\theta_e) +(1-k_e)\ln(1-\theta_e)
    \label{log_likelihood}
\end{equation}

Where $\theta_e = \alpha l_e^{\rho}$, and $k_e$ is the observed outcome of edge $e$. We note here that since $\alpha$ and $\rho$ are constrained to result in a valid probability calculated from $\alpha l_e^{\rho}$, the minimisation is subject to constraints and must satisfy the Karush-Kuhn-Tucker conditions\cite{KKT}. In practice, numerical optimisation of the log-likelihood in equation \ref{log_likelihood} was used to estimate $\alpha$ and $\rho$.

\subsection{Structural influence and network predictability}
\label{predict_method}
Depending on the values of the parameters for a given dataset, we might expect the observed values of $l_e$ to be predictive of subsequent change. Specifically, since $\rho$ controls the relationship between $l_e$ and the propensity for an edge to change, a high value of $\rho$ would suggest that $l_e$ would be more predictive of future change. Similarly for $\alpha$, within the constraints for $\alpha l_e^{\rho}$ to give the probability of an edge to change, a larger $\alpha$ factor will increase the distance between change probabilities for edges with different $l_e$, thus also strengthening the relationship between the value of $l_e$ and the propensity for an edge to change. In order to evaluate these effects, we make use of logistic regression for classification of edges into changing vs. unchanging from the values of $l_e$, and compare the results to a null model consisting of the average over multiple trials in which edges randomly change with probability equal to the fraction of observed changes. The data is split into training and test sets in a stratified manner, with 20\% used to test the model on unseen data. The predictions are compared according to balanced accuracy, defined as the average of recall obtained on each class, and Area Under Curve scores for both Receiver Operating Characteristic curves and Precision Recall curves.   

\section{Results}
\subsection{Validation of $l_e$ using toy networks}
\label{sec:le_validation}
\begin{figure}[H]
    \centering
    \includegraphics[width=0.9\textwidth]{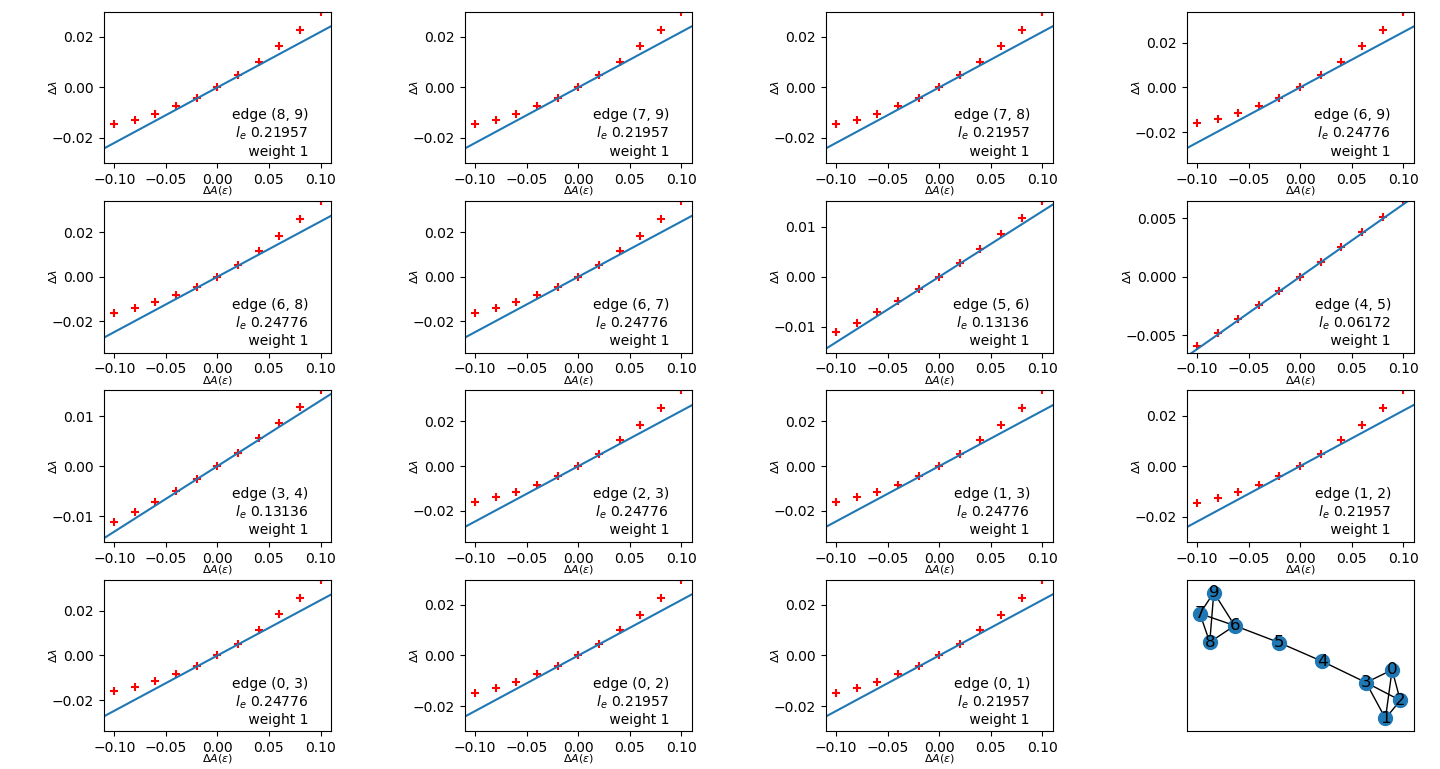}
    \caption{Scatter plot of perturbations $\Delta A$ and the resulting $\Delta \lambda$, compared to line of constant $l_e$. Barbell graph, with equal initial weights} 
    \label{fig:ind_edge_changes_barbell}
\end{figure}

\begin{figure}
    \centering
    \includegraphics[width=0.9\textwidth]{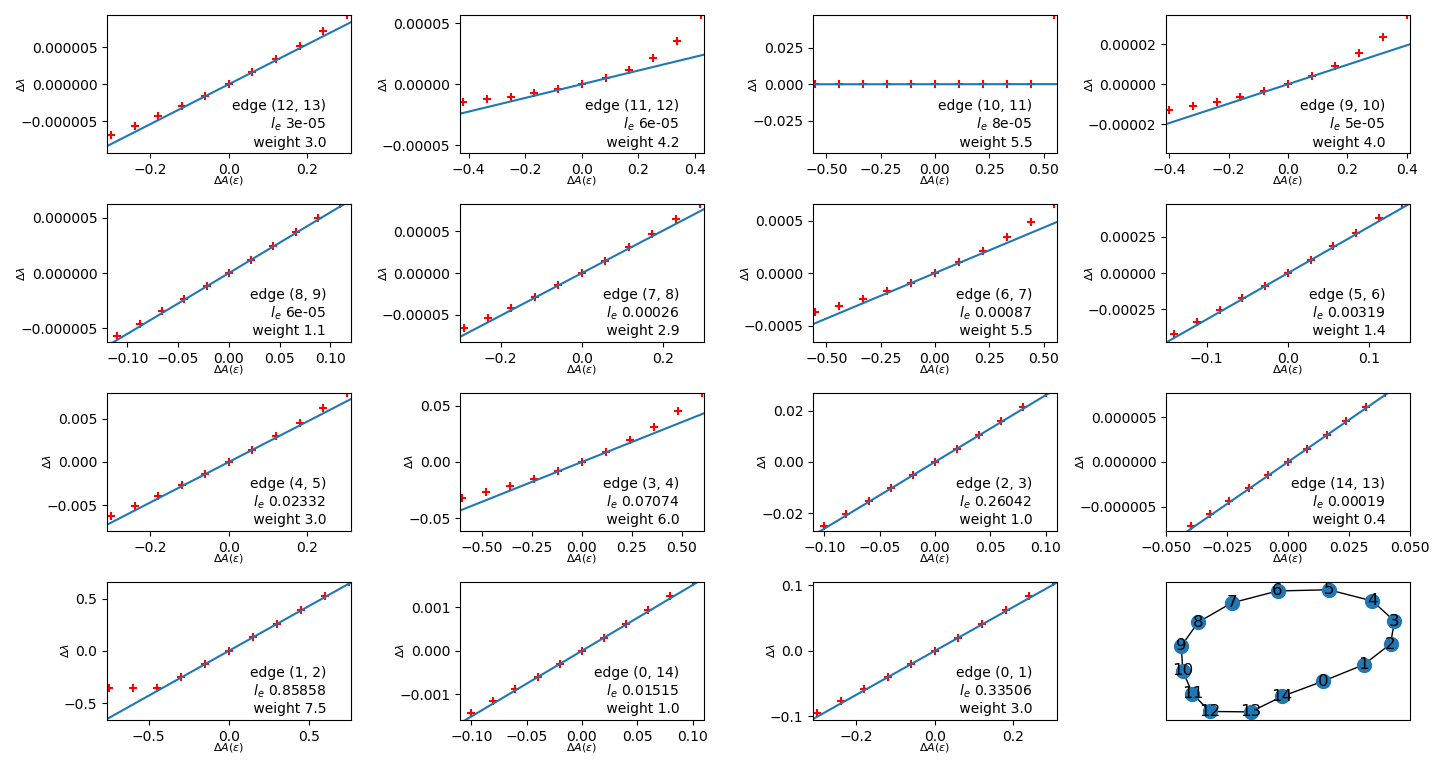}
    \caption{Scatter plot of perturbations $\Delta A$ and the resulting $\Delta \lambda$, compared to line of constant $l_e$. Ring graph with each edge independently assigned a random integer between 1 and 10.} 
    \label{fig:ind_edge_changes_ring}
\end{figure}

\begin{figure}
    \centering
    \includegraphics[width=0.9\textwidth]{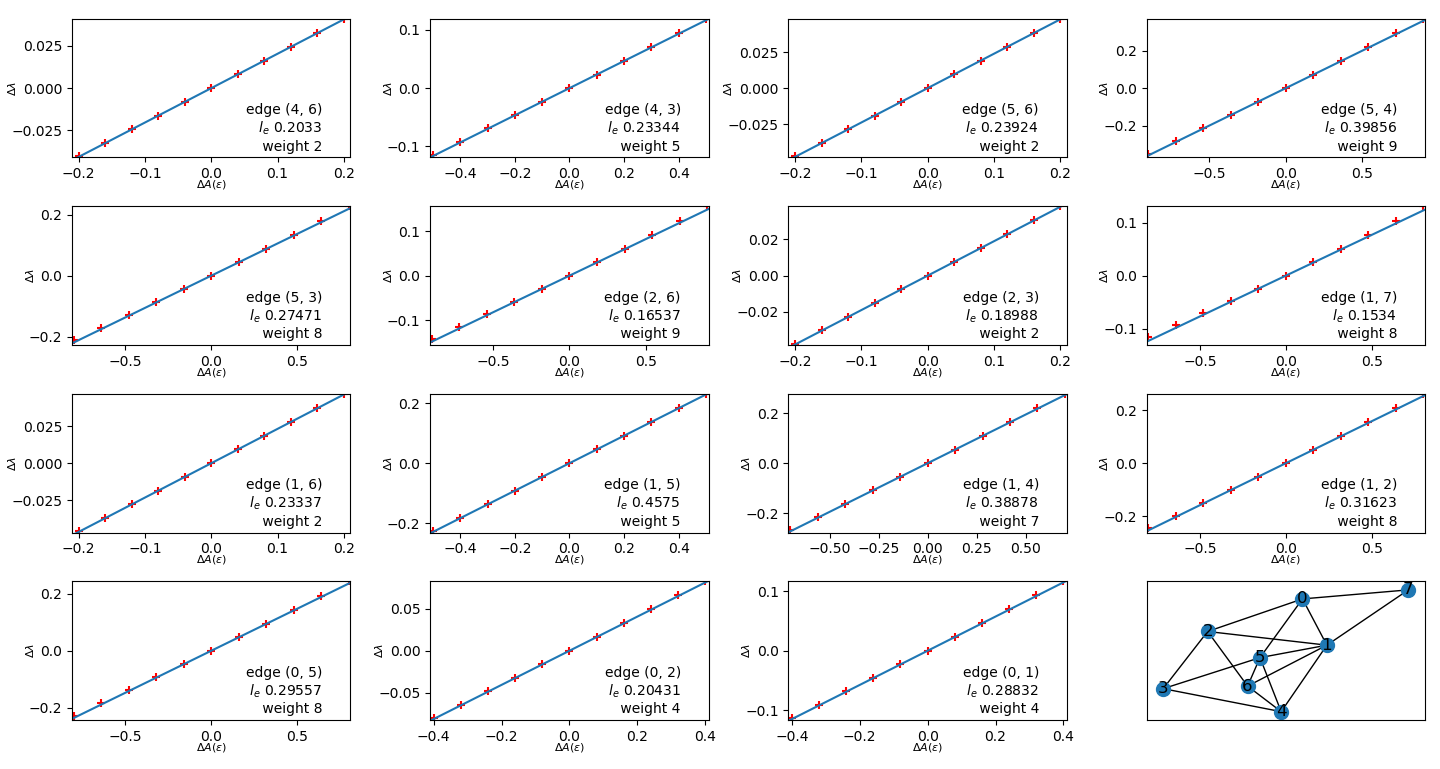}
    \caption{Scatter plot of perturbations $\Delta A$ and the resulting $\Delta \lambda$, compared to line of constant $l_e$. Erdős–Rényi graph with each edge independently assigned a random integer between 1 and 10.} 
    \label{fig:ind_edge_changes_er}
\end{figure}

Here we assess the extent to which the approximations made in calculating $l_e$ hold. We do this by approximating the change in eigenvalues as the coefficient weighted sum of the edge weight changes, $\Delta \lambda = \sum_{e}l_e \Delta A_e$, and comparing the gradient of this to the value of $l_e$. Our derivation of $l_e$ makes the simplification in assuming that edge changes occur independently of each other. Our first test thus considers the case of an individual edge changing at each timestep, and we consider perturbations applied to a barbell graph, to observe the effects of network structure, a ring graph, to observe the effects of weight with structural equivalence, and a Erdős–Rényi (ER) graph as a baseline. The results in figures \ref{fig:ind_edge_changes_barbell}, \ref{fig:ind_edge_changes_ring} \ref{fig:ind_edge_changes_er} show the line of constant $l_e$, overlaid with the observed $\Delta A_e$ and corresponding $\Delta \lambda$ values. 

 We see here that our linear approximation generally holds for relative edge changes less than $\Delta A = 0.05$. We also see for the barbell graph that $l_e$ captures the structural role of the edges, with edges in the cliques having higher values of $l_e$ than those in the bridge. For the ring graph, we observe a poorer fit for edges with low values of $l_e$, and the larger $l_e$ edges tend to be adjacent to edges with similar $l_e$ values. Although the edge with the largest weight also has the largest value of $l_e$, in general there does not appear to be a simple relationship between edge weight, or weight of neighbouring edges, and the value of $l_e$. For the weighted random network we see similar observations are made for the weighted ER graph, with the lowest $l_e$ values observed for more peripheral edges, and the two edges with the largest weights also having the highest $l_e$ values.
Further results for the case of a weighted barbell, and unweighted ring and random networks are shown in appendix \ref{appendix:perts}.

Results for the case of two edges changing are also shown in appendix \ref{appendix:perts}. In these we observe for the barbell graph better fit is observed for higher values of $l_e$. For the ring networks and random networks, we see that our model performs well if the observed edge has a larger value of $l_e$ than the other changing edge, but performs poorly when the value of $l_e$ is smaller. The case of complete structural equivalence and equal weights in the unweighted ring network shows good performance for all edges.

The breakdown of the method when there are multiple changes occurring between snapshots suggests that our approximation for $l_e$ may be better suited to a continuous or pseudo-continuous representation of a temporal network, which can be seen as the limit of a discrete temporal network in which each snapshot captures an individual edge change occurring at an infinitesimally different time to the neighbouring snapshot changes.

\subsection{Relationship between $l_e$ and the presence of edge changes}
\label{synthetic_exploration}
We can understand the role of the parameters $\alpha$ and $\rho$ by observing the effect of varying the parameters on the distributions of the values of $l_e$ for changing vs. non-changing edges, $P(\Delta A=0|\ln(l_e))$. We first consider this for data generated according to our model in equation \ref{markov_chain}, first keeping $\rho$ fixed and varying $\alpha$, then fixing $\alpha$ and varying $\rho$.

\subsubsection{Model with varying $\alpha$}
\begin{figure}[H]
    \centering
    \includegraphics[width=\textwidth]{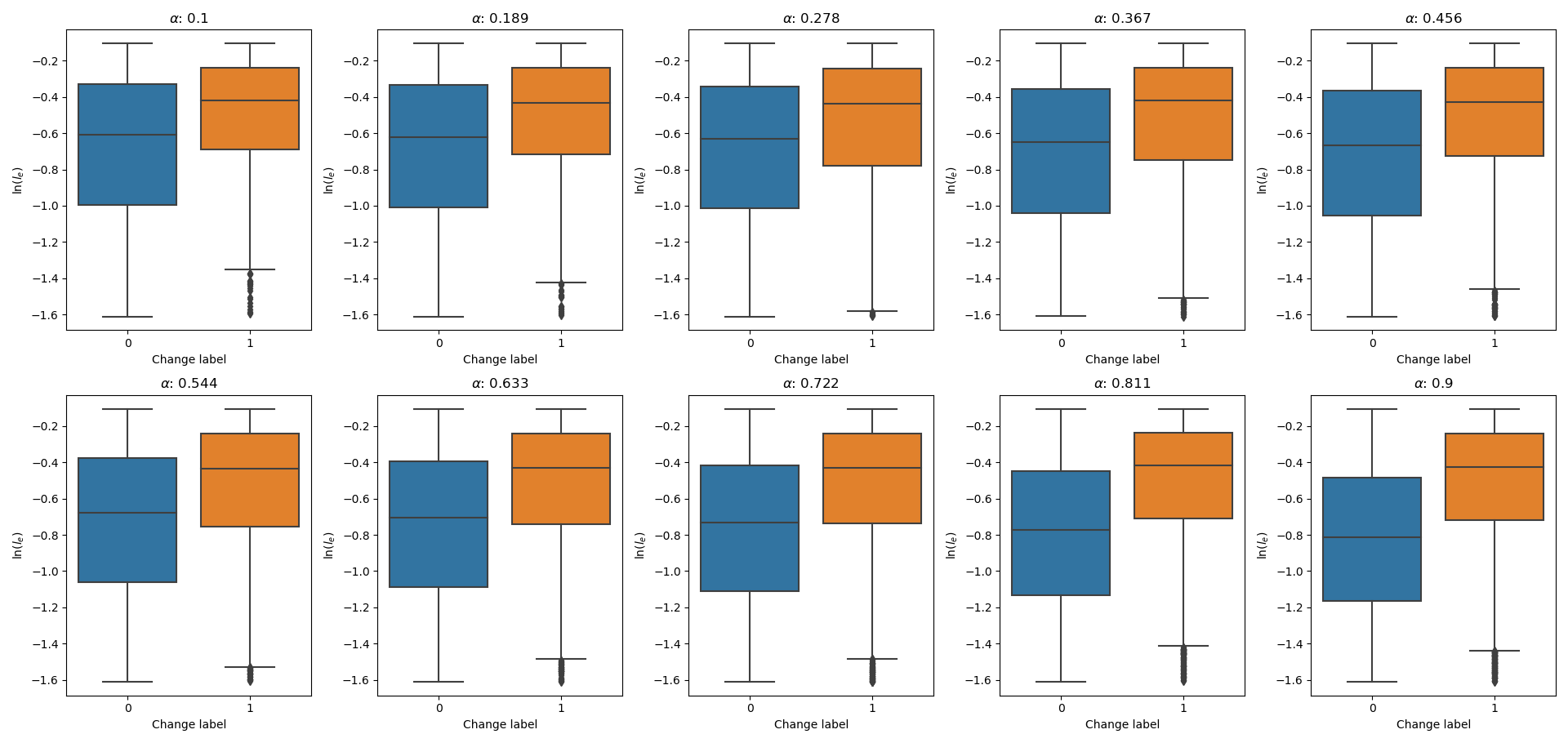}
    \caption{Distributions of $l_e$ for edge changes vs. no changes, when varying $\alpha$.}
    \label{fig:alpha_boxplots}
\end{figure}

\begin{figure}[H]
    \centering
    \includegraphics[width=\textwidth]{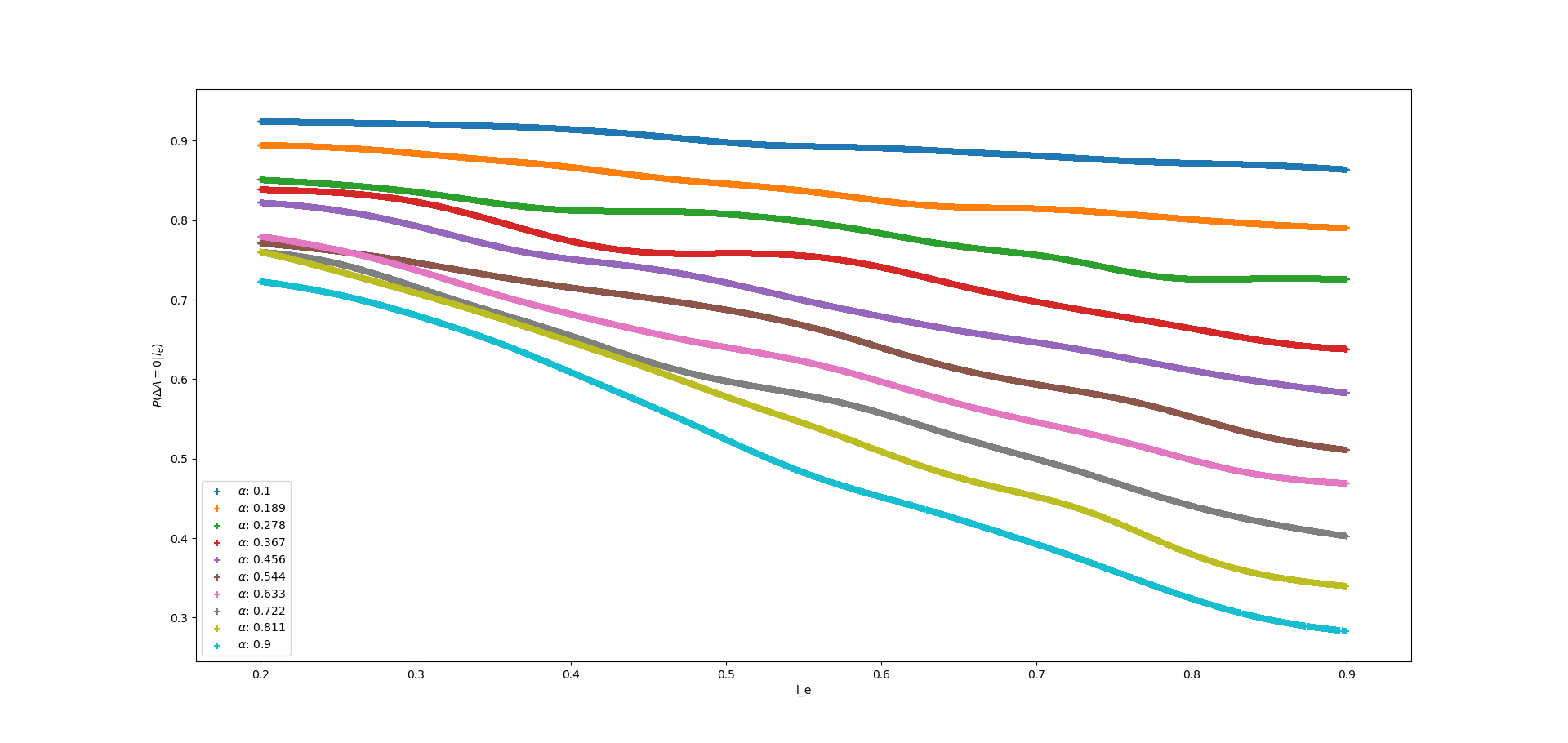}
    \caption{$P(\Delta A=0|\ln(l_e))$ as a function of $\ln(l_e)$ for $0.1<\alpha<1$}
    \label{fig:alpha_change_dist}
\end{figure}

Figures \ref{fig:alpha_boxplots} and \ref{fig:alpha_change_dist} show the resulting distributions for varying values of $\alpha$. We see that an increase in $\alpha$ results in a decrease in the probability of an edge to remain unchanged for all values of $l_e$, and for larger values of $\alpha$, the rate of increase of change probability with $l_e$ is slightly larger. 

\subsubsection{Model with varying $\rho$}
\begin{figure}[H]
    \includegraphics[width=\textwidth]{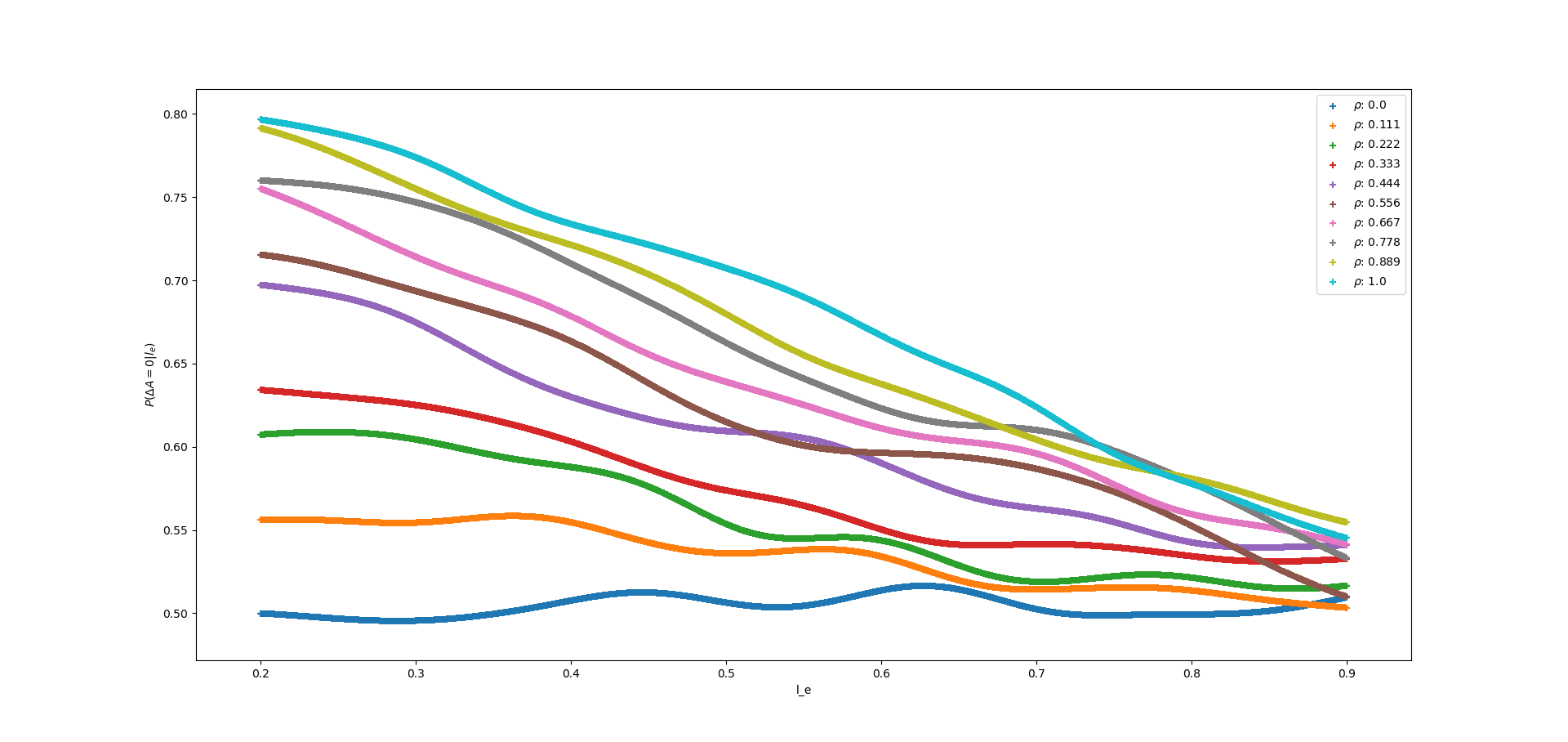}
    \caption{$P(\Delta A=0|\ln(l_e))$ as a function of $\ln(l_e)$ for $0<\rho<2.5$.} 
    \label{fig:da_0}
\end{figure}

Figure \ref{fig:da_0} shows the distributions of $P(\Delta A=0|\ln(l_e))$. We see here that for increasing $\rho$, the probability of observing no change increases, and also for increasing $l_e$, the probability decreases for a given $\rho$, at a rate that shows a significant dependence on $\rho$. 

\begin{figure}[H]
    \includegraphics[width=\textwidth]{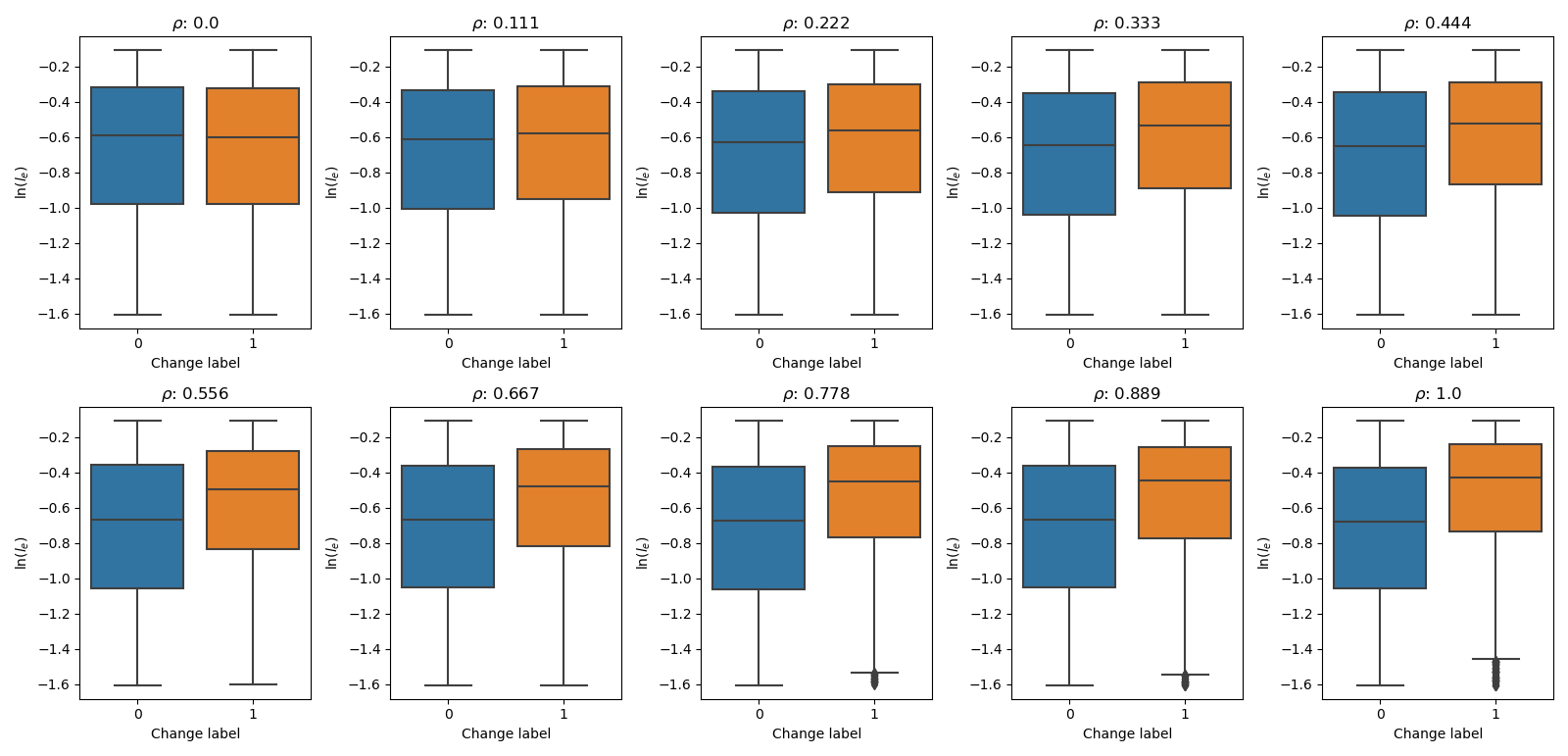}
    \caption{Distributions of $l_e$ value for the case of edge changes vs. no changes.} 
    \label{fig:da_0_box}
\end{figure}

\subsubsection{Predictability improvement with $\alpha$ and $\rho$}
\label{appendix:param_improvments}
\begin{figure*}
    \centering
\begin{subfigure}{.5\textwidth}
    \centering
    \includegraphics[width=\linewidth]{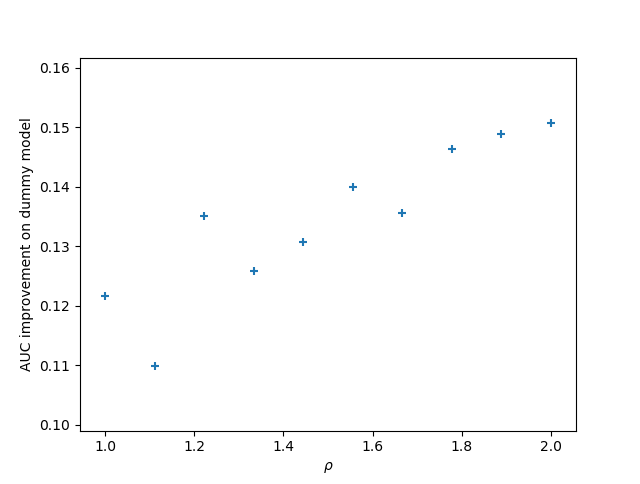}
    \caption{Model prediction performance improvement with $\rho$}
    \label{fig:rho_imp}
\end{subfigure}%
\begin{subfigure}{.5\textwidth}    
\centering
    \includegraphics[width=\linewidth]{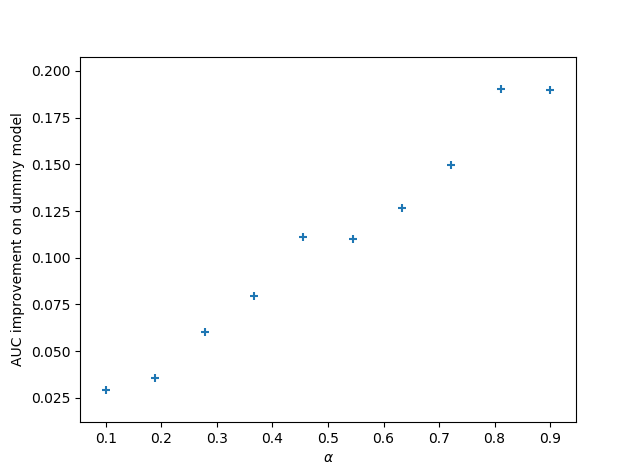}
    \caption{Model prediction performance with $\alpha$}
    \label{fig:alpha_imp}
\end{subfigure}
\caption{Improvement from dummy model of Precision-Recall AUC scores}

\end{figure*}

As detailed in section \ref{predict_method}, here we apply a logistic regression classifier with single feature $l_e$, to datasets with varying $\alpha$ and $\rho$.  Figures \ref{fig:rho_imp} and \ref{fig:alpha_imp} show the improvement in the test set Precision-Recall Area Under Curve scores for increasing values of each parameter. We see from these that increasing both parameters improves the predictability of changes given the value of $l_e$, consistent with our observations of the rate of increase of change probability being positively correlated with both $\alpha$ and $\rho$. 
\subsubsection{Static observations in real data}

We have seen in the above in application to synthetic networks that our model behaves as expected, with networks with a large $\rho$ (and $\alpha$) being more predictable. Now we explore the performance of our structural influence metric and model through the application to five real datasets. Firstly, given that our research has been motivated by a need to monitor risks in a financial setting, we considered a network of country level bilateral trade \cite{COW_data} and three different capital markets transaction datasets reported under MIFID II regulations. However, our methods can be applied more generally to any temporal networks, and due to the availability and high volume of research conducted into social networks (see \cite{Moreno}), we also considered a network of messages sent between College students \cite{college}. A full description of these can be found in appendix \ref{Appendix_datasets}.

In order to understand the usefulness of $l_e$ as a metric for structural importance, we first examine the edges that rank the highest according to their values of $l_e$ for the bilateral trade dataset, since the historical context of international trade can give us an idea of which edges we might expect to be `important'. For the bilateral trade dataset, we see the largest values of $l_e$ for the edge between Portugal and Spain in 1872, and considering the sum across all time, for Greece and Turkey. These are examples of edges with both nodes having large eigenvector centrality; edges involving only one central node are seen to have lower values of $l_e$. This means that inter-European edges almost exclusively make up the top 100 ranked edges, whereas the lowest ranked $l_e$ edges occur when one, or both, of the nodes have very low centrality scores. Similarly, for the other datasets, the highest values of $l_e$ were also observed for edges involving nodes with high eigenvector centrality. In general, we see that the rankings of $l_e$ are uncorrelated with the rankings of edges according to their betweenness centrality, or their mean value of $\Delta A$, however do for some cases
correlate with the product of the participating node's degrees and strengths.  
\begin{table}[H]
    \centering
    \begin{tabular}{|c|c|c|c|c|}
        \hline
       Dataset & Corr($l_e$, $\Delta A$) & Corr($l_e$, EBC) & Corr($l_e$, $deg_{n1}\times deg_{n2}$) &Corr($l_e$, $S_{n1}\times S_{n2}$)   \\
        \hline
       Bilateral Trade & -0.061 &  -0.397  & 0.352 & 0.786 \\
        College Messaging & -0.169 & 0.035 & 0.581 & 0.434\\
        Equity-1 & -0.104 & 0.135 & 0.717 & 0.320\\
        Equity-2 & -0.047 & 0.166 & 0.580 & 0.265\\
        Equity-3 & -0.010 & 0.041 & 0.923 & 0.763\\
        \hline
    \end{tabular}
    \caption{Spearman's rank correlations for $l_e$ with the rank by edge weight, edge betweenness centrality and product of nodes' degrees.}
    \label{tab:my_label}
\end{table}

\label{sec: res_static_real}

As these datasets contain large numbers of edges (the smallest contained 2785 edges), we cannot fully explore all of the individual observed values of $l_e$ as for the toy networks. Instead, we consider the probabilities of observing values of $l_e$ by making use of Kernel Density Estimation to estimate the probability density functions from the data. 

\begin{figure}
    \centering
    \hspace*{-1.8cm}
    \includegraphics[scale=0.4]{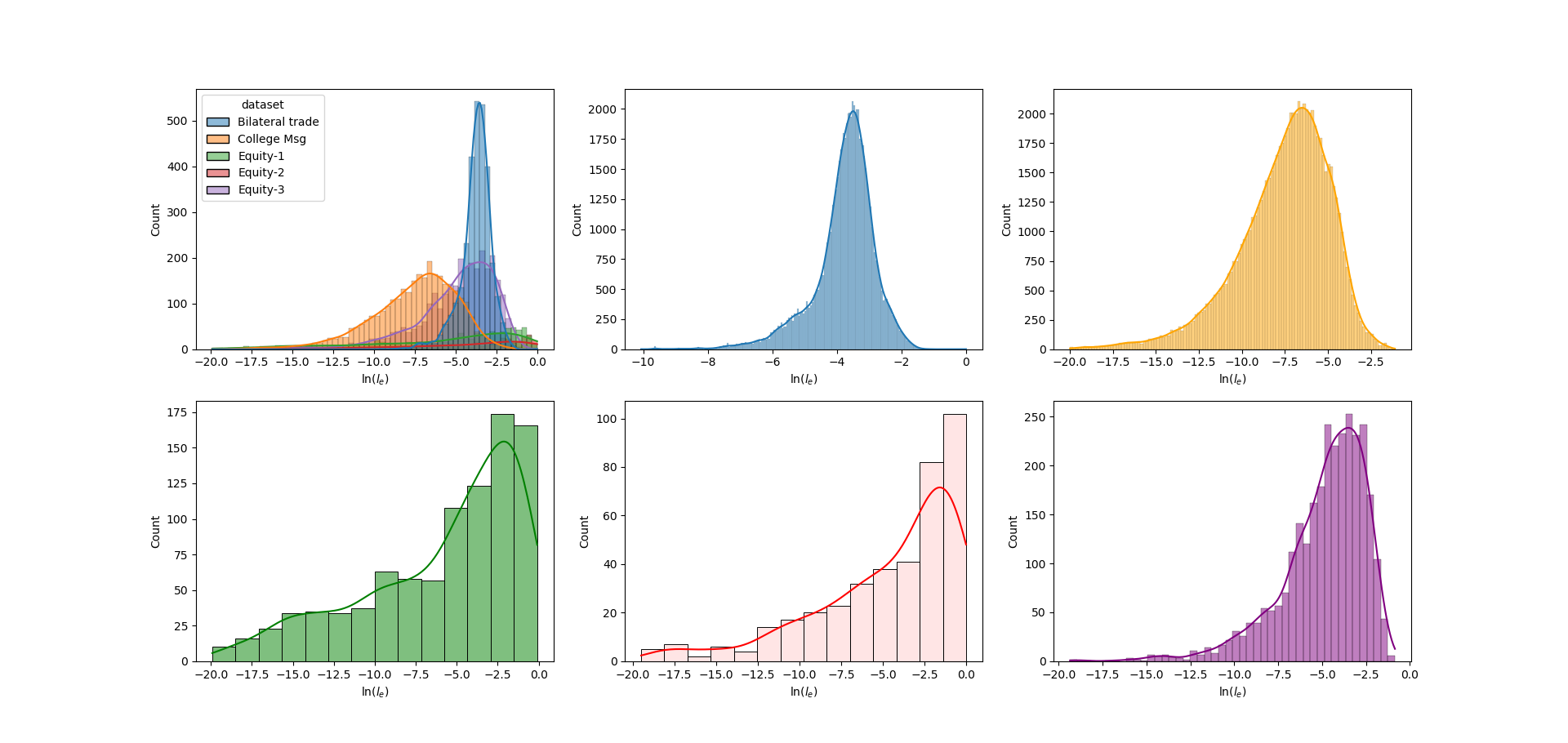}
    \caption{Probability distribution of the values of $\log(l_e)$ for different networks.} 
    \label{fig:p_l_e_dists}
\end{figure}

Figure \ref{fig:p_l_e_dists} shows the estimated Probability Density Functions of the logarithm of the value of $l_e$. We see from these that for all networks, the values observed for $l_e$ tend to be very small.  Omitting the tails of the distributions for diminishingly small values of $l_e$, we see a similarity in the values of $l_e$ observed across 3 similar equity datasets, and although across all 5 datasets analysed, the distribution is found to be approximately lognormal, the social network shows a much broader distribution of $l_e$. The peak of the distribution for the college messaging dataset is also much lower, observed at approximately $\ln(l_e)=-8.8$, whereas the bilateral trade dataset shows a peak at -3.3 , and the equity datasets at -3, -2.5 and -4.2. 

\subsubsection{Dynamic observations in real networks}
\label{real_nets_dyn}
\begin{figure}
    \centering
    \includegraphics[width = \textwidth]{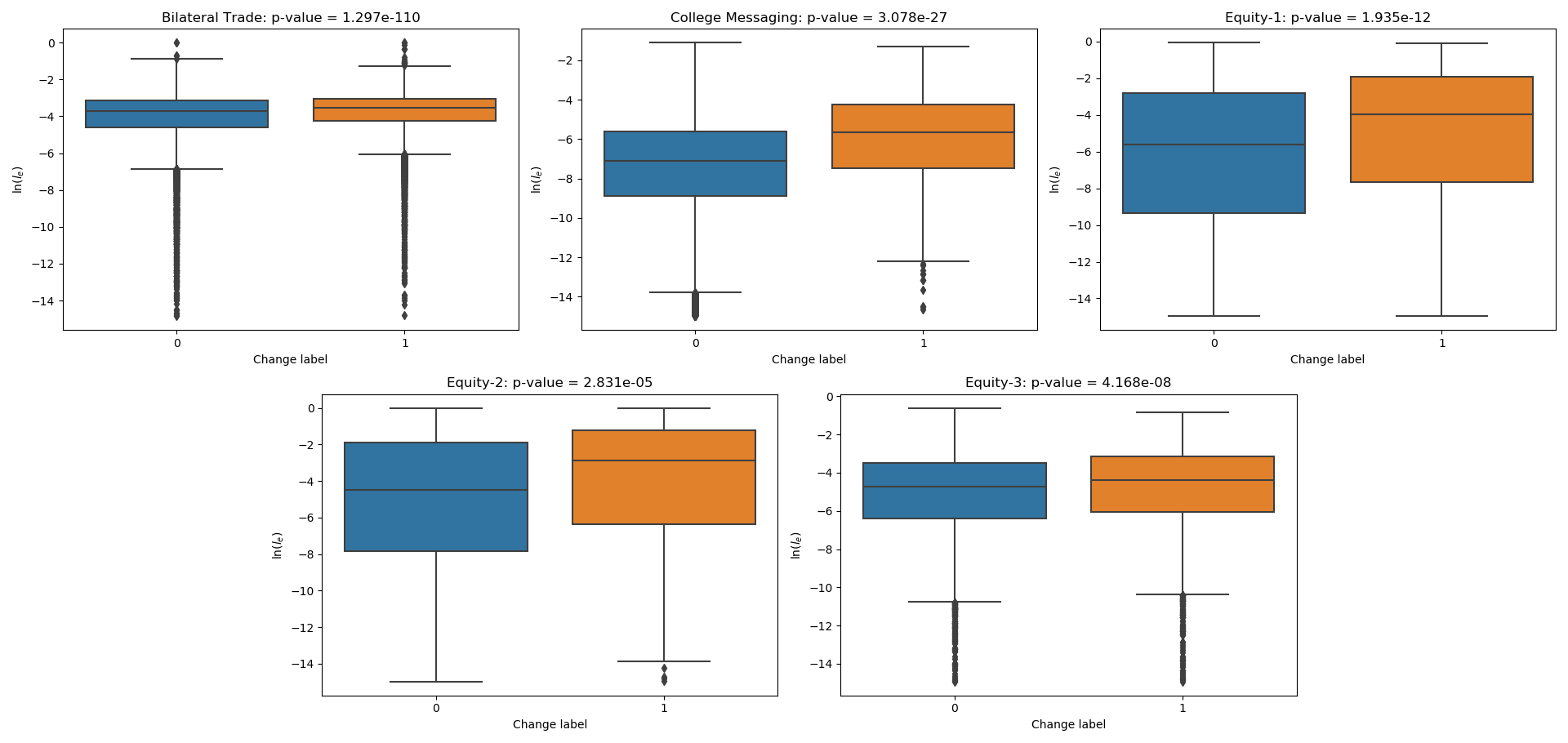}
    \caption{Boxplots showing the distribution of $l_e$ values observed according to the presence or absence of an edge subsequently changing.} 
    \label{fig:change_le}
\end{figure}

We now address the central concept of the relationship of $l_e$ observed for our real networks and the probability of an edge to change. 
Figure \ref{fig:change_le} shows the distributions of the $\ln(l_e)$ values observed for non-changing edges in comparison to changing edges. We see that in all cases, there is a shift in the mean value of $\ln(l_e)$ towards higher values for edges which do change, which would be suggestive of a positive $\rho$ parameter, and potentially the ability to predict the presence of changes given the value of $l_e$. The smallest shifts are observed for the Bilateral Trade dataset and Equity-3, which show negligible differences in the mean and quartiles of the values of $l_e$ for changes and no changes, suggesting that we might not expect predictability of changes from the values of $l_e$ in these cases. In all cases, the differences in the mean values of $l_e$ for change vs. no change is significant, with a two-sided t-test showing $p<0.05$ for all datasets.
\label{sec: res_dyn_real}

\begin{figure}[H]
    \centering
    \includegraphics[scale=0.5]{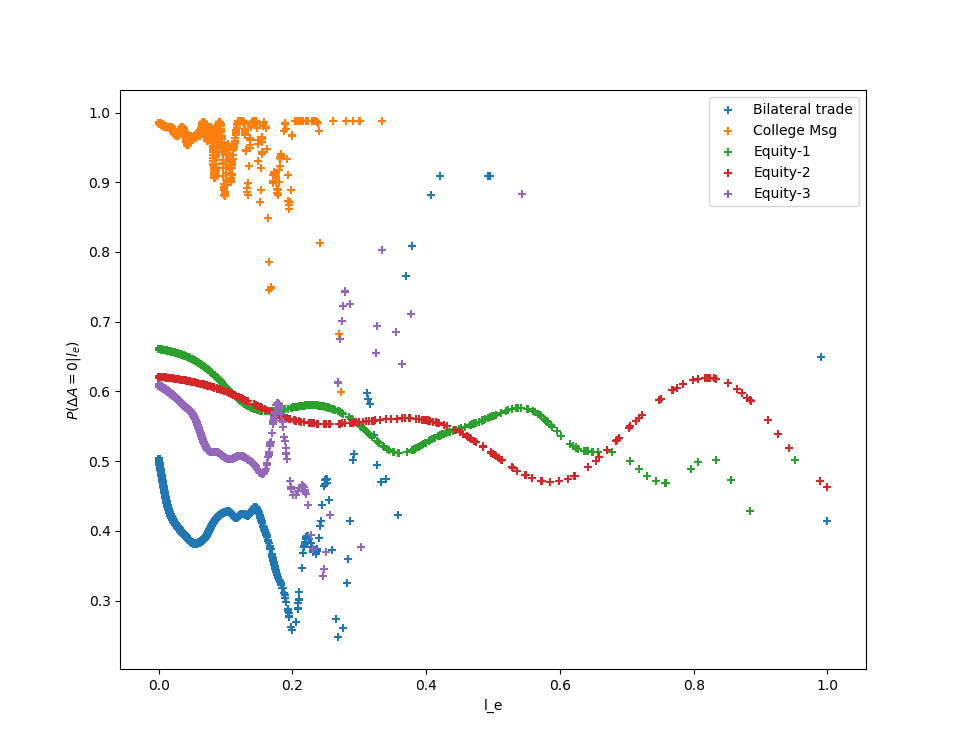}
    \caption{$P(\Delta A=0|\ln(l_e))$ as a function of $\ln(l_e)$ for the 5 real datasets.} 
    \label{fig:da_0_dists}
\end{figure}

To further understand how the value of $l_e$ relates to the probability for edges to change, we look at the distributions of $P(\Delta A=0|l_e)$ as shown in figure \ref{fig:da_0_dists}. Here we see a decreasing probability of $\Delta A=0$ for the bulk of the distribution for increasing $l_e$ for the bilateral trade and Equity-3 datasets, however the rarely observed edges with $l_e>0.3$ for these datasets show larger probabilities to remain unchanged. We again see a slight initial decrease for Equity-1 and 2 datasets however the relationship is clearly non-linear for large $l_e$. The college messaging dataset shows a much larger probability in general for edges to remain unchanged, and shows a very slight decrease in probability to remain unchanged for very small $l_e$ values, however is dominated by noise for $l_e >0.05$. 

 Referring back to section \ref{synthetic_exploration}, we considered the ideal cases of linear positive, neutral and negative relationships between $l_e$ and the probability of edge changes. In reality, as shown in figure \ref{fig:da_0_dists}, we see things are more complex, with different relationships apparent for different $l_e$ ranges. In particular, for edges with lower values of $l_e$, the negative relationship between the value of $l_e$ and the probability of an edge to remain unchanged suggests that a parameterisation of our model with positive value of $\rho$ would be effective in capturing the behaviour of the bulk of the network. However changes to the small handful of edges with the largest values of $l_e$ are less likely. These observations could suggest that there are a few structurally important edges which act to stabilise a system which would otherwise move towards a regime of instability.

\subsubsection{Estimation of $\alpha$ and $\rho$ from data}
In table \ref{grads_corrs}, we present the values of $\alpha$ and $\rho$ estimated for our 5 different datasets. The errors on these estimations are given by the inverse hessian of the Log-Likelihood, which is found by numerical approximation. In comparison with figures \ref{fig:alpha_change_dist} and \ref{fig:da_0_dists}, we see the ordering of the estimated value of $\alpha$ appears to agree with the positions of the college messaging dataset and the equity datasets. The parameter $\rho$ appears to correspond with the overall gradients observed in figure \ref{fig:da_0_dists} for the bulk of the distributions observed for low values of $l_e$. These observations suggest that our model is mostly capturing the imbalance of observed changes in the parameter $\rho$, and the overall average change probability for each dataset in the parameter $\alpha$.

\begin{table}[H]
\begin{center}
\begin{tabular}{ |c|c|c|}
\hline
 Dataset & Estimated $\alpha$ &Estimated $\rho$ \\
 \hline
Bilateral trade & $0.783 \pm 1.08\times 10^{-4}$ & $0.072 \pm 1.21\times 10^{-5}$  \\ 
College messaging & $0.033 \pm 3.03\times 10^{-6}$ & $0.270 \pm 2.76\times 10^{-6}$  \\ 
Equity-1 &  $0.392\pm 2.74\times 10^{-4}$ & $0.030\pm 2.82\times 10^{-5}$ \\   
Equity-2 & $0.401\pm 3.13\times 10^{-5}$ & $0.016\pm 4.93\times 10^{-5}$ \\
Equity-3 & $0.465 \pm 2.57 \times 10^{-4}$ & $0.036 \pm 4.21 \times 10^{-5}$ \\

 \hline
\end{tabular}
\end{center}
\caption{Estimated $\alpha$ and $\rho$ for the 5 real datasets}
 \label{grads_corrs}
 \end{table}

Figure \ref{fig:param_all} in the appendix shows the result of generating distributions of $P(\Delta A=0|l_e)$ for the estimated parameters, in comparison to the real datasets, restricted to $l_e<0.3$ in order to observe the bulk of the distributions. We see that the dataset generated according to the parameters estimated appear to show a reasonable agreement to the actual distribution, and differences here can be attributed to the differences in the initial network conditions. 

\subsubsection{Edge change predictability}
\begin{figure}[H]
    \centering
    \includegraphics[width=\textwidth]{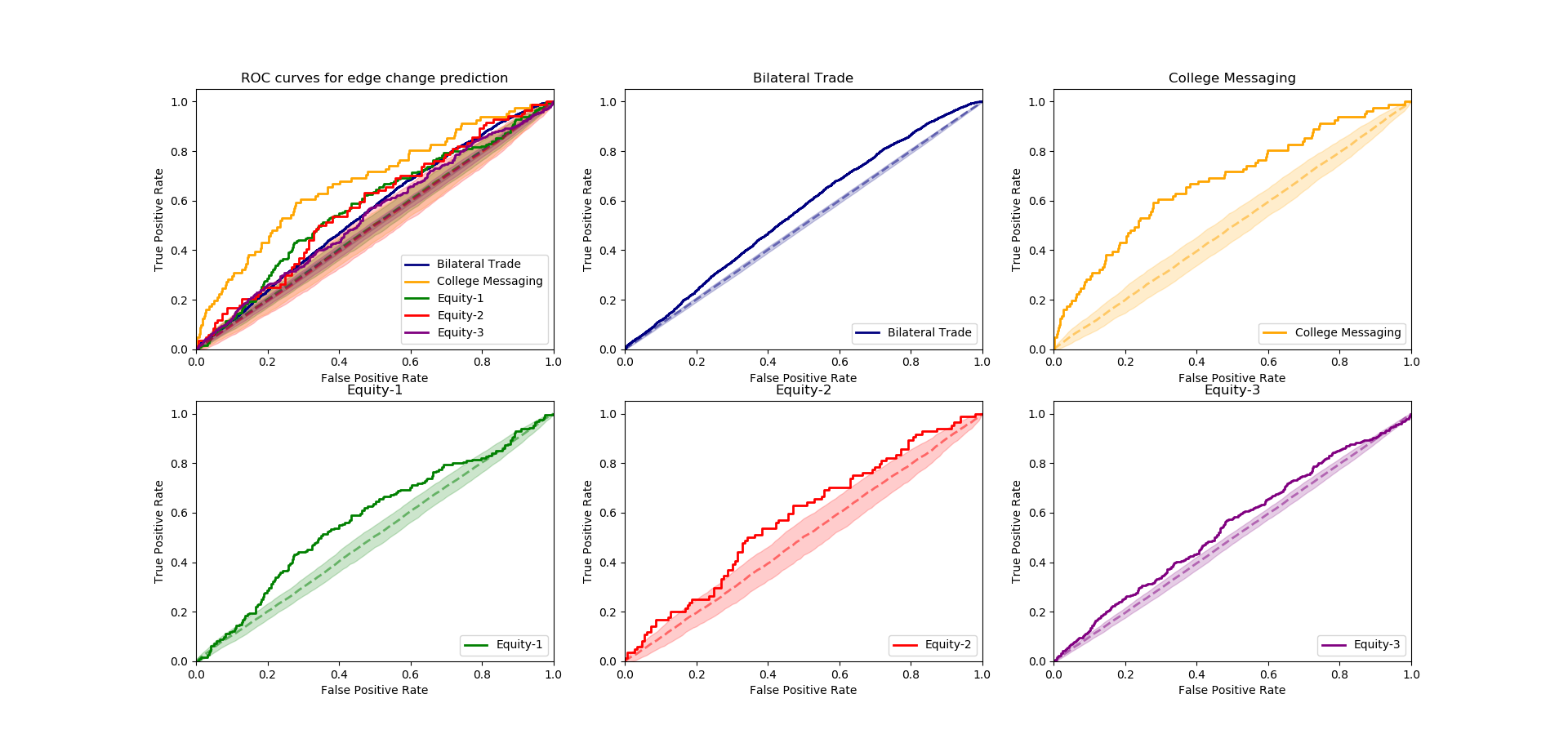}
    \caption{ROC curves for a logistic regression classifier making use of $\ln(l_e)$ to predict $\Delta A=1$. The dashed lines and shaded areas represent the mean 95\% confidence intervals for the dummy model.}
    \label{fig:ROC}
\end{figure}

\begin{figure}[H]
    \centering
    \includegraphics[width = \textwidth]{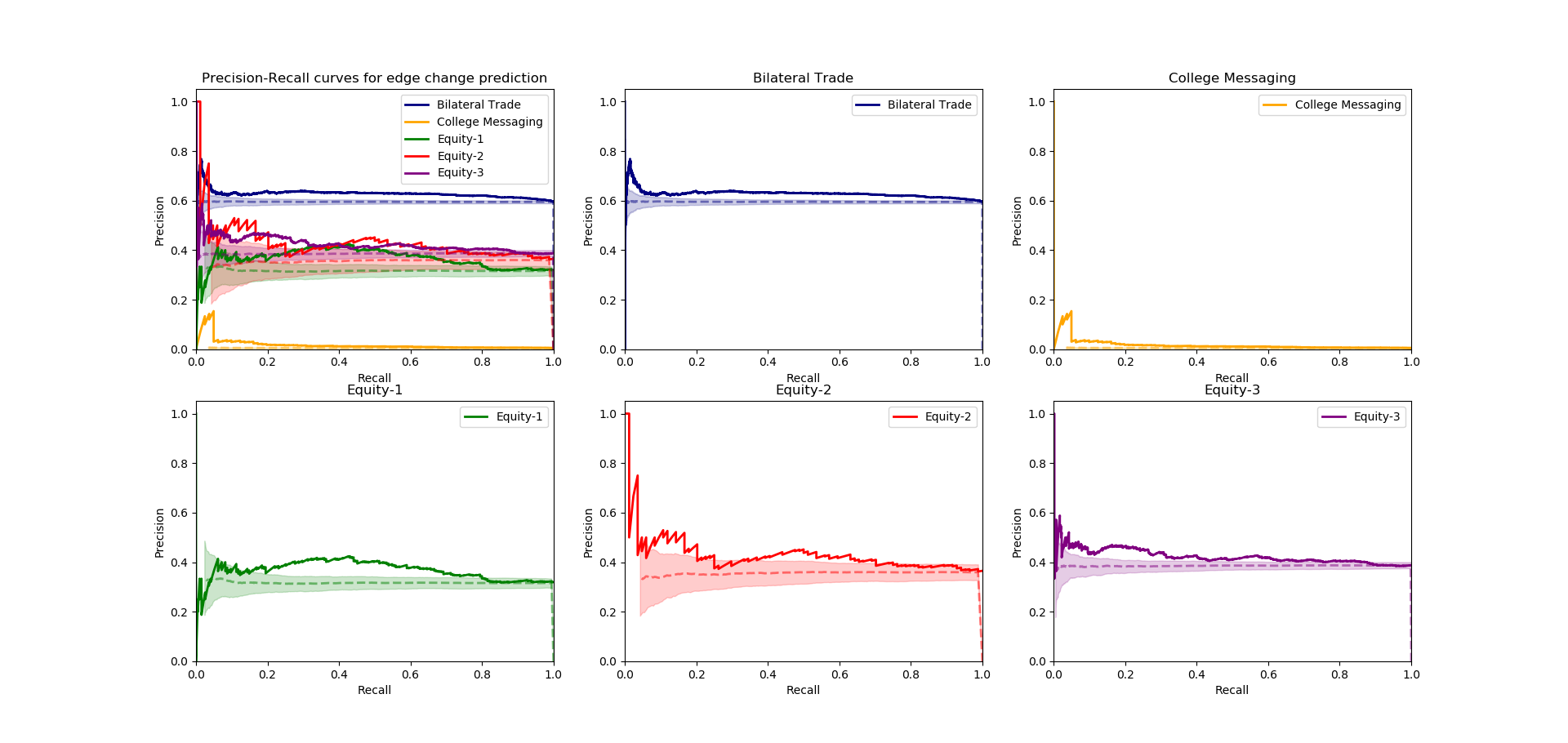}
    \caption{PR curves for a logistic regression classifier making use of $\ln(l_e)$ to predict $\Delta A=1$. The dashed lines represent the results for a stratified random allocation of labels.}
    \label{fig:PR}
\end{figure}

Given the non-zero estimated values of the parameters $\alpha$ and $\rho$, it is natural to assess the performance of using the value of $l_e$ to predict a subsequent change. Figures \ref{fig:ROC} and \ref{fig:PR} show the Receiver Operating Characteristic and Precision-Recall Curves for the 5 different datasets. 

All datasets are seen to perform slightly better than the dummy model, with better performance seen for the College Messaging dataset and Equity-1 and 2, which also show larger differences in the distribution of $l_e$ across change vs. no change in figure \ref{fig:da_0_box}. Poorer performance is seen for the bilateral trade and Equity-3 datasets, which show similar shaped distributions in figure \ref{fig:da_0_dists} with an initial steep decrease in probability to remain unchanged for increasing $l_e$, however this trend appears to reverse for $l_e>0.3$. These datasets also show little difference in the distribution of values observed in figure \ref{fig:da_0_box} and are found to have low values of $\rho$. Although the college messaging dataset shows the best performance, particularly in the left hand side of the ROC curve, this is driven by the significant class imbalance with only $5\%$ of the observations showing a non-zero $\Delta A$, as opposed to the bilateral trade dataset which shows a $20\%$ proportion of non-zero changes. 
\begin{table}[H]
\begin{center}
\begin{tabular}{ |c|c|c|c|}
\hline
 Dataset & Balanced accuracy & ROC AUC & Precision-Recall AUC\\
 \hline
Bilateral trade & 0.542 (0.5) & 0.554 (0.5) & 0.628 (0.595) \\ 
College messaging & 0.623 (0.5) & 0.678 (0.5) & 0.017 (0.005) \\ 
Equity-1 & 0.568 (0.5) & 0.576  (0.5) & 0.365 (0.313)	\\   
Equity-2 & 0.566 (0.5) & 0.579 (0.5) & 0.430 (0.351)	\\
Equity-3 & 0.527 (0.5) & 0.542 (0.5) & 0.424 (0.381)\\
 \hline
\end{tabular}
\end{center}
\caption{Values of Area Under Curve scores for ROC and Precision-Recall curves. Numbers in brackets represent the score achieved by a model which randomly predicts 1 or 0 in proportion to the dataset prior.}
 \label{rho_imp}
 \end{table}
 
\subsection{Relationship of between $l_e$ and size of weight changes}
\label{weight_params}
We now consider if the value of $l_e$ is observed to have an affect on the scale of subsequent edge changes. As in section \ref{synthetic_exploration}, we again consider data generated according to the model in equation \ref{markov_chain}, and we choose to take $U_{ij}^t = \mathcal{N}(\mu=0, \sigma=\beta l_e^\gamma)$. This introduces two new parameters, $\beta$ which controls the width of the distribution of edge changes, and $\gamma$ which controls the level to which $l_e$ influences the variance of the edge change distribution. 

\subsubsection{Variation of $\gamma$}
\begin{figure}[H]
    \centering
    \includegraphics[scale=0.5]{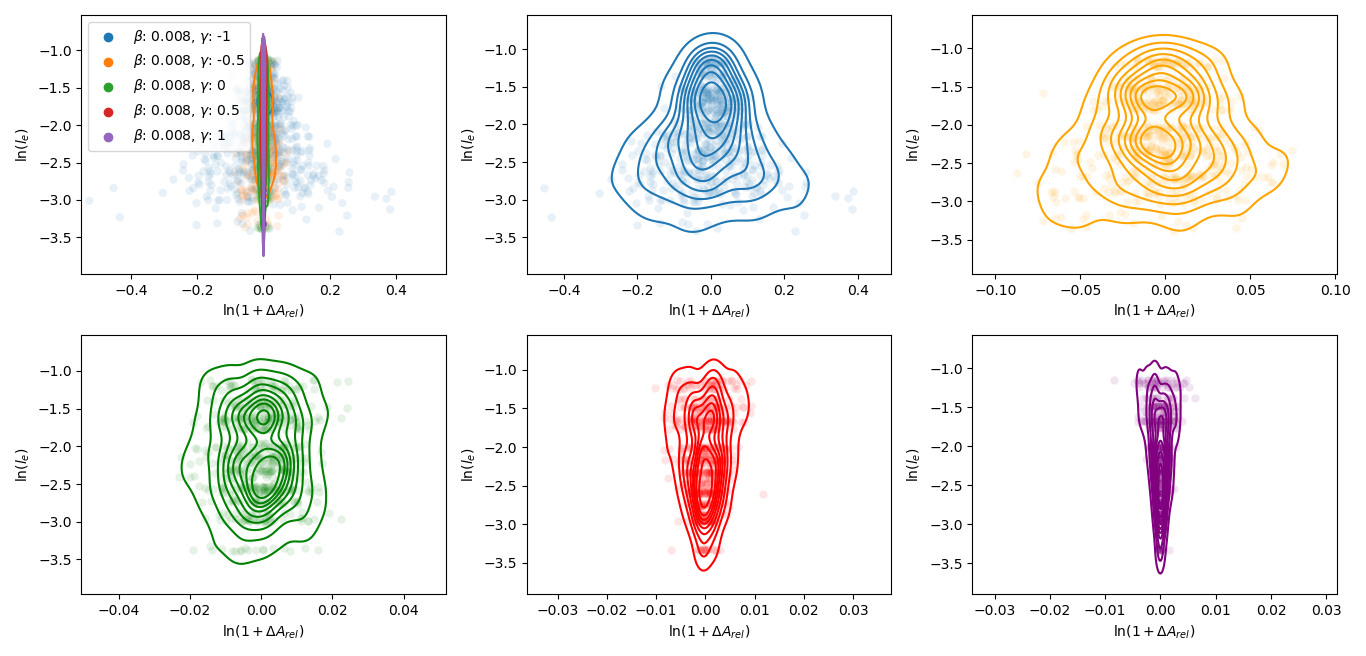}
    \caption{Distributions of $P(\ln(1+\Delta A),\ln(l_e)$ for fixed $\beta=0.008$, $-1<\gamma<1$} 
    \label{fig:gamma_var}
\end{figure}

Figure \ref{fig:gamma_var} shows the distributions of $P(\ln(1+\Delta A),
,l_e)$ for a range of values of $\gamma$. We see here that for positive $\gamma$, the width of the distribution widens for larger $l_e$. For negative $\gamma$, we see the opposite, that the width of the distribution becomes narrower for larger $l_e$.

\subsubsection{Variation of $\beta$}
\begin{figure}[H]
    \centering
    \includegraphics[scale=0.5]{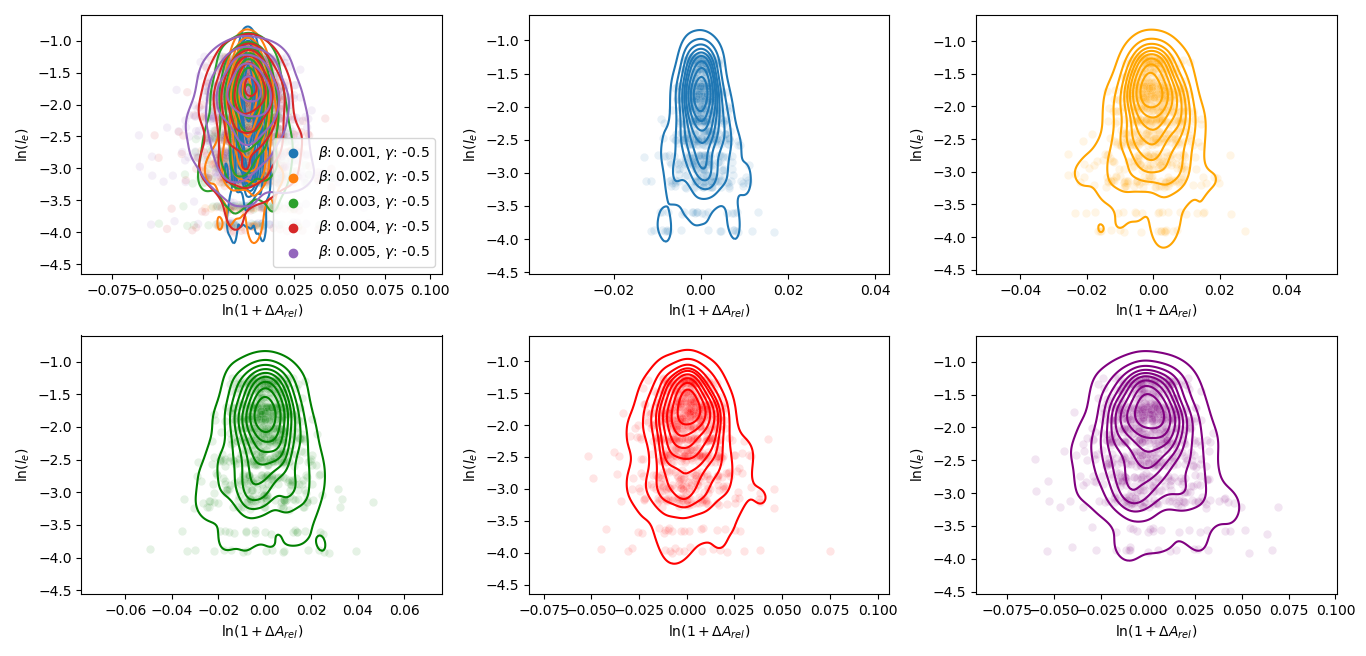}
    \caption{Distributions of $P(\ln(1+\Delta A),\ln(l_e)$ for fixed $\gamma=-0.5$, $0.001<\beta<0.005$.} 
    \label{fig:beta_var}
\end{figure}

Figure \ref{fig:beta_var} shows the distributions of $P(\ln(1+\Delta A), l_e)$ for a range of values of $\beta$. We see here that as $\beta$ increases, the width of the distributions increase. 
\subsubsection{Weight distributions for real networks}
\begin{figure}
    \centering
    \includegraphics[width=\textwidth]{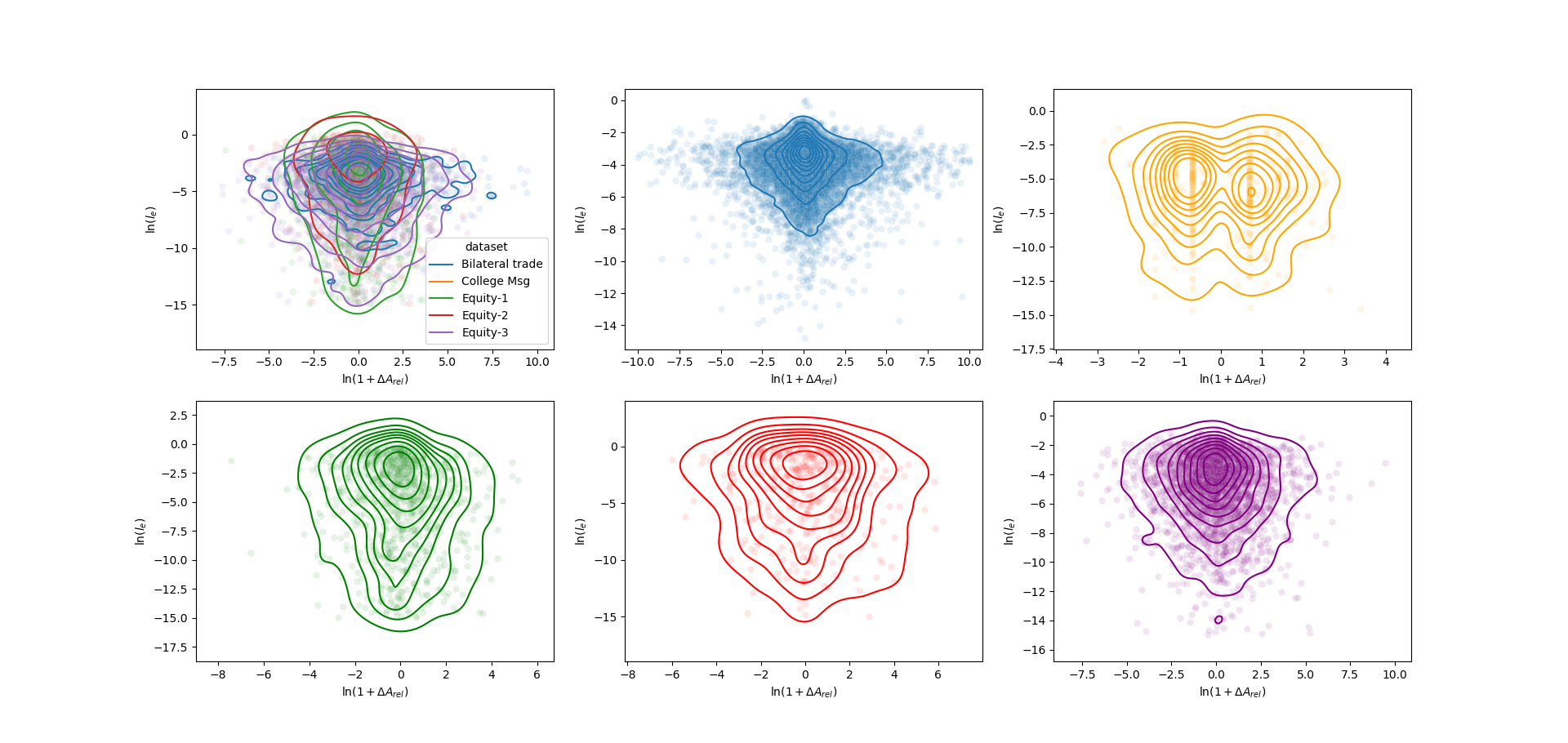}
    \caption{Contours showing the distributions of $P(\ln(1+\Delta A),\ln(l_e)$ for the 5 real datasets. Underlying observations of $\ln(l_e)$ and $\ln(1+\Delta A_{rel})$ represented by the dots underlying these.}
    \label{fig:heatmaps_real}
\end{figure}

We now consider the same 5 real datasets considered in section \ref{real_nets_dyn}. Figure \ref{fig:heatmaps_real} shows the distributions of $P(\ln(1+\Delta A),\ln(l_e))$ for the case of edges that do change, i.e. $\Delta A\neq 0$ for the five real networks. Here $\Delta A$ refers to the relative change in the value of the edge weight from $t_0$ to $t_1$, which takes values in the interval $[-1, \infty]$, and $l_e$ is measured at time $t_0$. Infinite values for $\Delta A$, corresponding to the case of a new edge appearing, were observed but are not captured in the plots. The prominence of these across the different datasets are $4.7\%$ of the bilateral trade dataset, $0.086\%$ of the college messaging dataset, $0.012\%$, $0\%$ and $0.0028\%$ of the equity datasets\footnote{The prominence of new edges has been significantly reduced by focusing on the giant component}. We see a slight widening of the distributions for larger values of $l_e$ for Equity-1 and 2 datasets, and to a larger extent for the third equity dataset. The bilateral trade dataset shows initial widening as $l_e$ increases, however narrows again for the largest $l_e$ edges. The college messaging dataset shows two distinct peaks, corresponding to changes in edge weight of $\pm 1$, which are over-represented in this dataset as it is unweighted, and the edge weight solely represents the count of interactions in the time window of consideration. The slight widening for larger $l_e$ for all datasets is suggestive of a positive relationship between the value of $l_e$ and the variance of the distribution of subsequent edge changes.  

\subsubsection{Parameter estimation for $\beta$ and $\gamma$}

\begin{table}[H]
\begin{center}
\begin{tabular}{ |c|c|c|}
\hline
 Dataset & Estimated $\beta$ & Estimated
 $\gamma$\\
 \hline
Bilateral trade & $1.11\times 10^{-4} \pm 8.93\times 10^{-18}$ & $1.93\pm 1.10\times 10^{-27} $\\ 
College messaging & $7.77\times 10^{-5} \pm 1.10\times 10^{-16}$ & $1.15 \pm 3.07\times 10^{-27}$ \\ 
Equity-1 & $1.51\times 10 ^{-5}\pm 4.55\times 10^{-24}$ & $1.35\pm 2.55\times 10^{-36}$ \\   
Equity-2 & $1.42\times 10^{-4} \pm 1.36\times 10^{-26}$ & $1.32 \pm 5.81\times 10^{-37}$\\
Equity-3 &  $1.63\times 10^{-4} \pm 2.20\times 10^{-26}$ & $1.42\pm 1.43\times 10^{-36}$\\
 \hline
\end{tabular}
\end{center}
\caption{Estimated $\beta$ and $\gamma$ for the 5 real different datasets}
 \label{beta_gamma_ests}
 \end{table}
All 5 datasets show positive values of $\gamma$, suggestive of a relationship between the width of the distribution of edge changes and the value of $l_e$. The dataset with the highest values for $\gamma$, the bilateral trade dataset dataset, also shows the largest level of bias towards larger change distribution width for higher $l_e$ in figure \ref{fig:heatmaps_real}. Correspondingly, the lowest $\gamma$ value is seen for the college messaging dataset, which shows the least bias towards larger changes occurring for larger values of $l_e$. The values for $\beta$ are similar across the 5 datasets, and all relatively low. It is difficult to draw conclusions from these, as the behaviours controlled by the two parameters cannot be separated and observed alone in the distributions in figure \ref{fig:heatmaps_real}.
 
\section{Discussion \& Conclusion}
The ability to understand how microscopic changes in networks affect the macroscopic evolution across time is one of the key challenges in dynamic network analysis. In this study we have begun to explore the use of derivatives of network spectra to capture this. We derive a measure of edge based structural influence, $l_e$, and explore the extent to which the value is indicative of future changes. We first of all demonstrated that for small and isolated perturbations applied to the network, the eigenvalue derivative is approximated well by equation \ref{initial theory}. However, we observe the approximation breaks down for multiple changes happening during the same time snapshot, suggesting that the measure may be more suited to a continuous or pseudo-continuous representation of network evolution, in which each time snapshot contains a single edge change. 

Considering the 5 real datasets, we observe lognormal distributions of the values of $l_e$, indicating structural influence dominated by a small handful of edges. 
We propose a model in which the probability for an edge to change is given by  $\alpha l_e^{\rho}$. This model allows us to control the extent to which $l_e$ dictates the propensity for an edge to change, and also controls the scale of a subsequent change. Focusing on the former, we observe similarities in the shapes of the distributions of $P(\Delta A=0,\ln(l_e))$ when generating synthetic networks according to this model and those observed in the data, and the values observed for $\alpha$ and $\rho$ are suggestive of a relationship between the value of $l_e$ and the subsequent presence of change. In using $l_e$ in a logistic regression classifier to predict change, we see that $l_e$ is slightly predictive of change in all cases, but only marginally so for the case of the bilateral trade and Equity-3 datasets. This corresponds with our observations of small values of $\rho$ for these datasets, along with similar, non-linear distributions shapes for the probability of no change for increasing $l_e$. These observations indicate that the static structural importance can be indicative of the presence of a subsequent change, however more work is needed to understand the shape of the distribution and the identification of different $l_e$ regimes. We will also consider taking a similar approach with other measures of edge importance, for example edge gravity \cite{Helander}. More work is also needed to understand the subsequent impact on the global network structure of an edge changing. It may be that a change influential edge could act to destabilise a system; conversely, the change could move the system towards a state of stability. We will look to investigate this in future analyses.

We note here that $\alpha$ and $\rho$ themselves are useful parameters that could be used to classify networks according to their growth stability. A large value of $\alpha$ would be an indicator for larger levels of overall network activity. A network with very large $\rho$ would be characterised by changes occurring to the nodes with the largest $l_e$, conversely, a network with very small $\rho$ would see changes distributed across all edges, regardless of the value of $l_e$. In the context of financial markets, these contrasting situations would require different approaches, and $\rho$ could be used by policy makers to inform which asset classes should be monitored as a whole (for the case of small $\rho$) or following an approach targeting those nodes with the highest $l_e$. 

Our model doesn't account for edges appearing and disappearing in the network, and assumes that edge changes are independent of each other. For the first limitation, we note that edge appearance and disappearance would be unlikely to heavily influence the behaviour of the Equity networks, as we observed very low percentages ($0.012\%$, $0\%$ and $0.0028\%$) of new edges appearing \footnote{This is largely due to the preprocessing steps applied to the network, since we are only considering the giant component.
}, but for the other two networks this behaviour is much more prominent at 4.7\% for the bilateral trade network and 0.086\% for the college messaging network.  The measure $l_e$ itself is able to assess the importance of an edge that subsequently disappears, and also those that appear between two existing nodes, so understanding how these appearances and disappearances can be captured in a model for network growth would be highly beneficial for future work. On the second point, we noted in our exploration of toy networks that the ability of our approximation of the eigenvalue derivative breaks down for multiple edge changes present. Conversely, many works such as Bandi et. al. have noted that predictability is aggregation scale specific. In future work we will thus investigate the trade-off between improved approximation of $l_e$ for the quasi-continuous limit in which each time snapshot contains a single edge change, and improved predictability for larger aggregation scales. In addition to this, further analysis is needed to assess the effectiveness of $l_e$ as an indicator for risk, as so far we understand that the value of $l_e$ bears some relationship to how the network subsequently changes, but we have not yet considered the resultant changes of edges with high values of $l_e$, and how these have an effect on the rest of the network in terms of risk and stability. This is another area we will pursue in future work. We will also consider extending our methods to consider structural node importance, which is of use to policy makers who may wish to monitor which players could have an adverse impact on markets. It is also worth noting that although using raw transaction data gives us the lowest granularity view of the data, our work has so far not considered the higher order effects of trading behaviour on price. Such an effect results in the influence of edges reaching disconnected components, which cannot be captured by our methods, so we will consider generalising our methods to allow for networks with disconnected components. Finally, we will consider using our methods for classification of a large number of networks, and also extend our methods to understand the parameters which control the resultant weight changes. 

\section{Data availability}
The datasets referred to as Equity-1, Equity-2 and Equity-3 in this paper, were extracted from a dataset of transaction reports collected by the FCA under MIFID II regulations. The datasets were used under agreement from the data owners at the Financial Conduct Authority for the current study, and are not publicly available. A data note describing the data is shown in appendix \ref{Appendix_datasets}, and comparable analysis is conducted for all investigations for two open source datasets (see below). 

The dataset referred to as Bilateral Trade are hosted by Katherine Barbieri, University of South Carolina, and Omar Keshk, Ohio State University, available at http://correlatesofwar.org.
The dataset referred to as College Messaging are available in the Stanford Large Network Dataset Collection repository, https://snap.stanford.edu/data/CollegeMsg.html 

An implementation of the methods referenced in this paper can be found at \cite{my_code}.

\bibliographystyle{unsrt}
\bibliography{thesisbib.bib}

\begin{thebibliography}{10}

\bibitem{Nier}
E.~Nier, J.~Yang, T.~Yorulmazer, and A.~Alentorn.
\newblock Network models and financial stability.
\newblock {\em Bank of England Working Paper no. 346}, 2008.

\bibitem{Silva}
T.C. Silva, S.~M. Guerra, B.~M. Tabak, and R.C. de~Castro~Miranda.
\newblock Financial networks, bank efficiency and risk-taking.
\newblock {\em Banco Central do Brasil Working Paper n. 428.}, 2016.

\bibitem{Allen}
H.~J. Allen.
\newblock Financial stability regulation as indirect investor/consumer
  protection regulation: Implications for regulatory mandates and structure.
\newblock {\em Legal Studies Research Paper Series Research Paper 17-5}, 2017.

\bibitem{Haldane}
A.~Haldane and R.~May.
\newblock Systemic risk in banking ecosystems.
\newblock {\em Nature 469, 351–355 https://doi.org/10.1038/nature09659},
  2011.

\bibitem{Ren}
X.~Ren and L.~Lu.
\newblock Review of ranking nodes in complex networks.
\newblock {\em Chinese Science Bulletin 59. 1175. 10.1360/972013-1280.}, 2014.

\bibitem{Stegehuis}
C.~Stegehuis, R.~van~der Hofstad, and van Leeuwaarden.
\newblock Epidemic spreading on complex networks with community structures.
\newblock {\em Sci Rep 6, 29748 https://doi.org/10.1038/srep29748}, 2016.

\bibitem{Ren1}
Guangming Ren and Xingyuan Wang.
\newblock Epidemic spreading in time-varying community networks.
\newblock {\em Chaos: An Interdisciplinary Journal of Nonlinear Science},
  24(2):023116, 2014.

\bibitem{Lai}
Lai Y.C., Motter A.E., and Nishikawa T.
\newblock Attacks and cascades in complex networks.
\newblock {\em In: Ben-Naim E., Frauenfelder H., Toroczkai Z. (eds) Complex
  Networks. Lecture Notes in Physics, vol 650. Springer, Berlin, Heidelberg.
  https://doi.org/10.1007/978-3-540-44485-5 14}, 2004.

\bibitem{Xuhui}
H.~Xu, J.~Zhang, J~Yang, and L.~Lun.
\newblock Identifying important nodes in complex networks based on
  multiattribute evaluation.
\newblock {\em Mathematical Problems in Engineering, vol. Article ID 8268436,
  https://doi.org/10.1155/2018/8268436}, 2018.

\bibitem{Moreno}
Y.~Moreno, M.~Nekovee, and A.~F. Pacheco.
\newblock Dynamics of rumor spreading in complex networks.
\newblock {\em Phys. Rev. E}, 69:066130, Jun 2004.

\bibitem{Restrepo}
J.G. Restrepo, E.~Ott, and B.R. Hunt.
\newblock Weighted percolation on directed networks.
\newblock {\em hys. Rev. E71,036151}, 2005.

\bibitem{Bollobas}
B.~Bollobas, C.~Borgs, J.~Chayes, and O.~Riordan.
\newblock Percolation on dense graph sequences.
\newblock {\em hys. Rev. E71,036151}, 2005.

\bibitem{Wang}
Y.~Wang, Z.~Di, and Y.~Fan.
\newblock Identifying and characterizing nodes important to community structure
  using the spectrum of the graph.
\newblock {\em PLoS ONE 6(11):e27418. doi:10.1371/journal.pone.0027418}, 2011.

\bibitem{Stanley}
E.~Stanley, L.~Lü, L.~Pan, T.~Zhou, and Y.~Zhang.
\newblock Toward link predictability of complex networks.
\newblock {\em 112 (8) 2325-2330; DOI: 10.1073/pnas.1424644112}, 2015.

\bibitem{Battiston}
S.~Battiston, M.~Puliga, R.~Kaushik, P.~Tasca, and G.~Caldarelli.
\newblock Debtrank: Too central to fail?financialnetworks, the fed and systemic
  risk.
\newblock {\em Sci. Rep.2,541; DOI:10.1038/srep00541}, 2012.

\bibitem{Barucca}
P.~Barucca and F.~Lillo.
\newblock The organization of the interbank network and how ecb unconventional
  measures affected the e-mid overnight market.
\newblock {\em Comput Manag Sci (2017), arxiv: 1511.08068}, 2017.

\bibitem{Helander}
M.E. Helander and S.~McAllister.
\newblock The gravity of an edge.
\newblock {\em Appl Netw Sci 3, 7, https://doi.org/10.1007/s41109-018-0063-6},
  2018.

\bibitem{Yu}
E.~Yu, D.~Chen, and J.~Zhao.
\newblock Identifying critical edges in complex networks.
\newblock {\em Sci Rep 8, 14469 https://doi.org/10.1038/s41598-018-32631-8},
  2018.

\bibitem{Wang_eigs}
Y.~Wang, D.~Chakrabarti, C.~Wang, and C.~Faloutsos.
\newblock Epidemic spreading in real networks: An eigenvalue viewpoint.
\newblock {\em 22nd International Symposium on Reliable Distributed Systems.
  Proceedings.}, 2003.

\bibitem{Pei}
S.~Pei and H.A. Makse.
\newblock Spreading dynamics in complex networks.
\newblock {\em Journal of Statistical Mechanics: Theory and Experiment}, 2013.

\bibitem{Bardoscia}
M.~Bardoscia, G.~Bianconi, and G.~Ferrara.
\newblock Multiplex network analysis of the uk otc derivatives market.
\newblock {\em Bank of England Working Papers}, 2018.

\bibitem{May}
R.~M. May.
\newblock Will a large complex system be stable?
\newblock {\em NATURE VOL. 238}, 1972.

\bibitem{Restoy}
F.~Restoy.
\newblock Market integration: the role of regulation.
\newblock {\em Speech by Chairman of the Financial Stability Institute, Bank
  for International Settlements, at the IIF Market fragmentation roundtable,
  Washington DC, United States, 10 April 2019.}, 2019.

\bibitem{Samuelson}
P.~A. Samuelson.
\newblock General proof that diversification pays.
\newblock {\em The Journal of Financial and Quantitative Analysis, 2(1):pp.
  1–13}, 1967.

\bibitem{Bardoscia1}
M.~Bardoscia, S.~Battiston, F.~Caccioli, and G.~Caldarelli.
\newblock Pathways towards instability in financial networks.
\newblock {\em Nature Communications 8:14416 | DOI: 10.1038/ncomms14416}, 2016.

\bibitem{Markose}
S.~Markose, S.~Giansante, GatkowskiM., and Shaghaghi~A. R.
\newblock Too interconnected to fail: Financial contagion and systemic risk in
  network model of cds and other credit enhancement obligations of us banks.
\newblock {\em COMISEF WORKING PAPERS SERIESWPS-033}, 2010.

\bibitem{Caccioli1}
F.~Caccioli, M.~Marsili, and P.~Vivo.
\newblock Eroding market stability by proliferation of financialinstruments.
\newblock {\em arXiv:0910.0064 [q-fin.TR]}, 2009.

\bibitem{Tobin}
J.~Tobin.
\newblock The new economics one decade older.
\newblock {\em The Eliot Janeway Lectures on Historical Economics in Honour of
  Joseph Schumpeter 1972 (Princeton University Press, Princeton, US);Eastern
  Economic Journal IV153-159}, 1978.

\bibitem{Bianconi}
G.~Bianconi, T.~Galla, M.~Marsili, and P.~Pin.
\newblock Effects of tobin taxes in minority game markets.
\newblock {\em J. Econ. Behavior and Organization,Volume 70, Issues 1–2},
  2009.

\bibitem{Brock}
W.A. Brock, C.H. Hommes, and Wagener F.O.O.
\newblock More hedging instruments may destabilize markets.
\newblock {\em Journal of Economic Dynamics and Control 33 1912–1928}, 2009.

\bibitem{Aste1}
T.~Aste, W.~Shaw, and T.~Di~Matteo.
\newblock Correlation structure and dynamics in volatile markets.
\newblock {\em New Journal of Physics}, 2010.

\bibitem{Mazzarisi}
P.~Mazzarisi, P.~Barucca, F.~Lillo, and D.~Tantari.
\newblock A dynamic network model with persistent links and node-specific
  latent variables, with an application to the interbank market.
\newblock {\em European Journal of Operational Research Volume 281, Issue 1,
  16, Pages 50-65}, 2020.

\bibitem{Barucca1}
P.~Barucca and F.~Lillo.
\newblock Disentangling bipartite and core-periphery structure in financial
  markets.
\newblock {\em Chaos, solitons and fractals, arxiv: 1551.08830}, 2015.

\bibitem{Barucca3}
B.~Barucca, M.~Bardoscia, F.~Caccioli, M.~D'Errico, G.~Visentin, S.~Battiston,
  and G.~Caldarelli.
\newblock Network valuation in financial systems.
\newblock {\em Mathematical Finance.30:1181–1204.}, 2020.

\bibitem{Sora}
K.~Soram\"{a}ki, M.~Bech, J.~Arnold, R.~Glass, and W.~Beyeler.
\newblock The topology of interbank payment flows.
\newblock {\em Physica A 379 (2007) 317–333}, 2007.

\bibitem{Barabasi}
A.~Barabasi and R.~Albert.
\newblock Emergence of scaling in random networks.
\newblock {\em arXiv:cond-mat/9910332 [cond-mat.dis-nn]}, 1999.

\bibitem{Falkenberg}
M.~Falkenberg, J.~Lee, S.~Amano, K.~Ogawa, K.~Yano, Y.~Miyake, T.~Evans, and
  K.~Christensen.
\newblock Identifying time dependence in network growth.
\newblock {\em PHYSICAL REVIEW RESEARCH 2, 023352}, 2020.

\bibitem{Bazzi}
M.~Bazzi, L.~Jeub, A.~Arenas, S.~Howison, and M.~Porter.
\newblock A framework for the construction of generative models for mesoscale
  structure in multilayer networks.
\newblock {\em arXiv:1608.06196v5 [cs.SI]}, 2019.

\bibitem{Peixoto}
T.~P. Peixoto and M.~Rosvall.
\newblock Modelling sequences and temporal networks with dynamic community
  structures.
\newblock {\em Nat Commun 8, 582, https://doi.org/10.1038/s41467-017-00148-9},
  2017.

\bibitem{Watts}
D.~Watts and S.~Strogatz.
\newblock Collective dynamics of ‘small-world’ networks.
\newblock {\em Nature 393, 440–442 https://doi.org/10.1038/30918}, 1998.

\bibitem{Bianconi1}
Barabási~A.L. Bianconi~G.
\newblock Competition and multiscaling in evolving networks.
\newblock {\em Europhysics Letters. 54 (4): 436–442. arXiv:cond-mat/0011029},
  2001.

\bibitem{Kobayashi}
T.~Kobayashi and M.~Génois.
\newblock Two types of densification scaling in the evolution of temporal
  networks.
\newblock {\em Physical Review E 102, 052302 2020}, 2020.

\bibitem{Holme}
P.~Holme and J.~Saramaki.
\newblock Temporal networks.
\newblock {\em Phys. Rep. 519, 97-125 (2012). arXiv:1108.1780 [nlin.AO]}, 2010.

\bibitem{Tang}
J.~Tang, M.~Musolesi, C.~Mascolo, V.~Latora, and V.~Nicosia.
\newblock Analysing information flows and key mediators through temporal
  centrality metrics.
\newblock {\em SNS '10: Proceedings of the 3rd Workshop on Social Network
  Systems Article No.: 3 Pages 1–6 https://doi.org/10.1145/1852658.1852661},
  2010.

\bibitem{KKT}
H.~W. Kuhn and A.~W. Tucker.
\newblock Nonlinear programming.
\newblock {\em Proceedings of the Seconds Berkeley Symposium on Mathematical
  Statistics and Probability, pp. 481-492}, 1950.

\bibitem{COW_data}
K.~Barbieri and Omar K.
\newblock Correlates of war project trade data set codebook, version 4.0, 2016.

\bibitem{college}
P.~Panzarasa, T.~Opsahl, and K.~M. Carley.
\newblock Patterns and dynamics of users' behavior and interaction: Network
  analysis of an online community., 2009.

\bibitem{my_code}
I.~Seabrook.
\newblock Structural edge importance - python implementation.
\newblock {\em Github repository}, 2020.

\end{thebibliography}
\appendix
\section{Results of perturbing two nodes in toy networks}
\label{appendix:perts}
\begin{figure}[H]
    \centering
    \includegraphics[width = \textwidth]{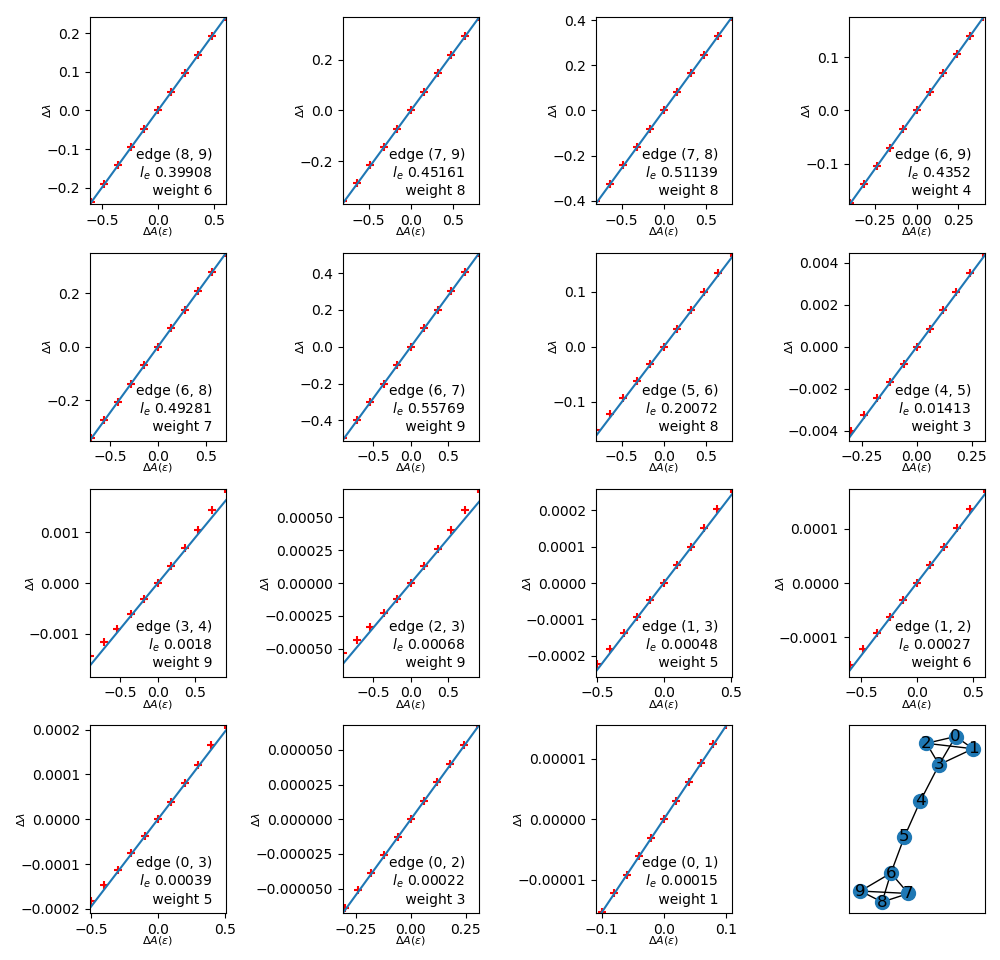}
    \caption{Scatter plot of perturbations $\Delta A$ and the resulting $\Delta \lambda$, compared to line of constant $l_e$. Weighted barbell network} 
    \label{fig:ind_edge_changes_barbell_w}
\end{figure}

\begin{figure}[H]
    \centering
    \includegraphics[width = \textwidth]{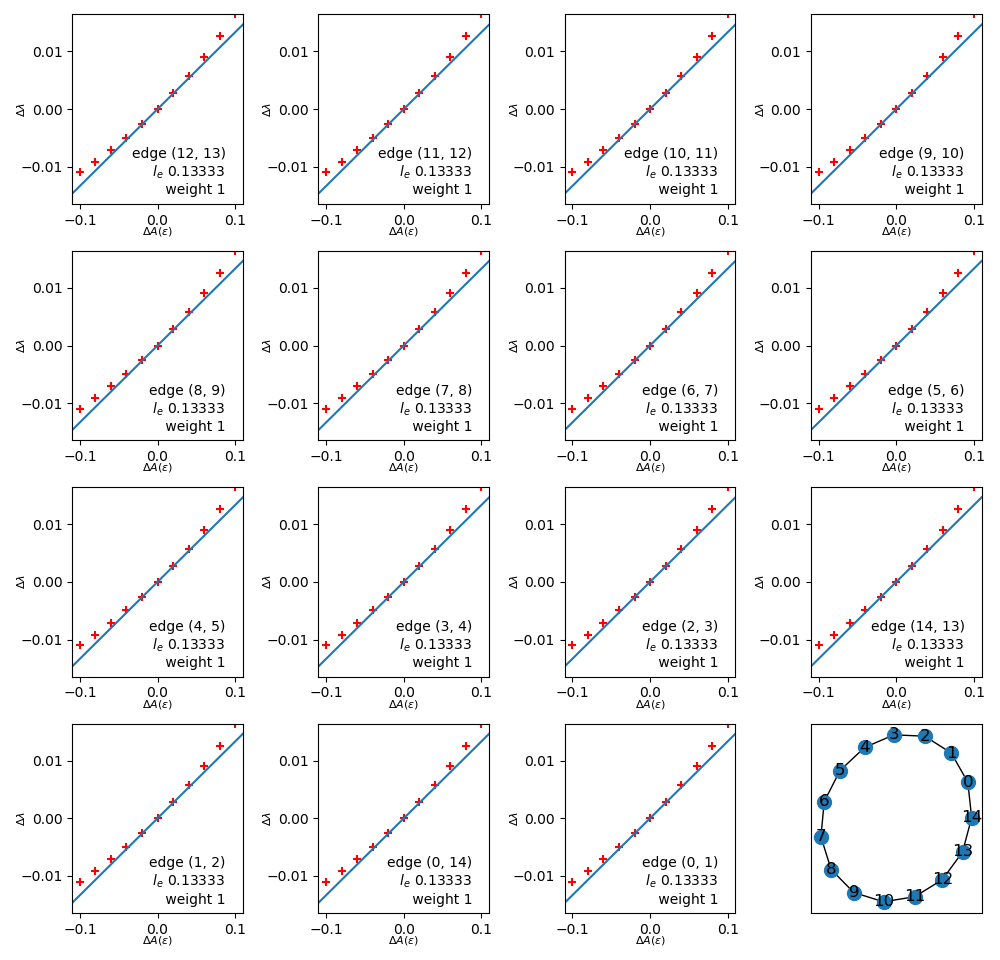}
    \caption{Scatter plot of perturbations $\Delta A$ and the resulting $\Delta \lambda$, compared to line of constant $l_e$. Unweighted ring network.} 
    \label{fig:ind_edge_changes_ring_u}
\end{figure}

\begin{figure}[H]
    \centering
    \includegraphics[width = \textwidth]{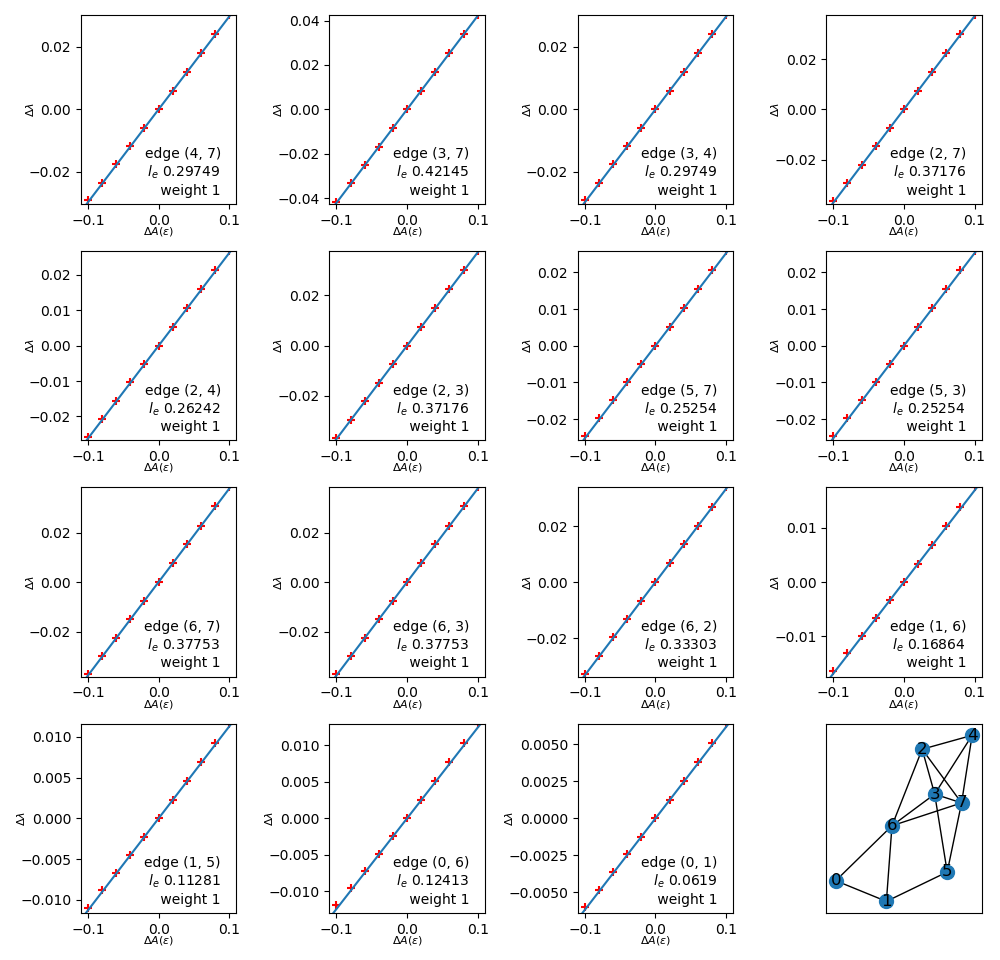}
    \caption{Scatter plot of perturbations $\Delta A$ and the resulting $\Delta \lambda$, compared to line of constant $l_e$. Unweighted Erdős–Rényi network} 
    \label{fig:ind_edge_changes_er_u}
\end{figure}

The plots shown in figures \ref{fig:ind_edge_changes_barbell_w} - \ref{fig:ind_edge_changes_er_u} show additional results for the case of perturbing single edges for weighted barbell networks, and unweighted ring and Erdős–Rényi networks. 

\begin{figure}[H]
    \centering
    \includegraphics[width = \textwidth]{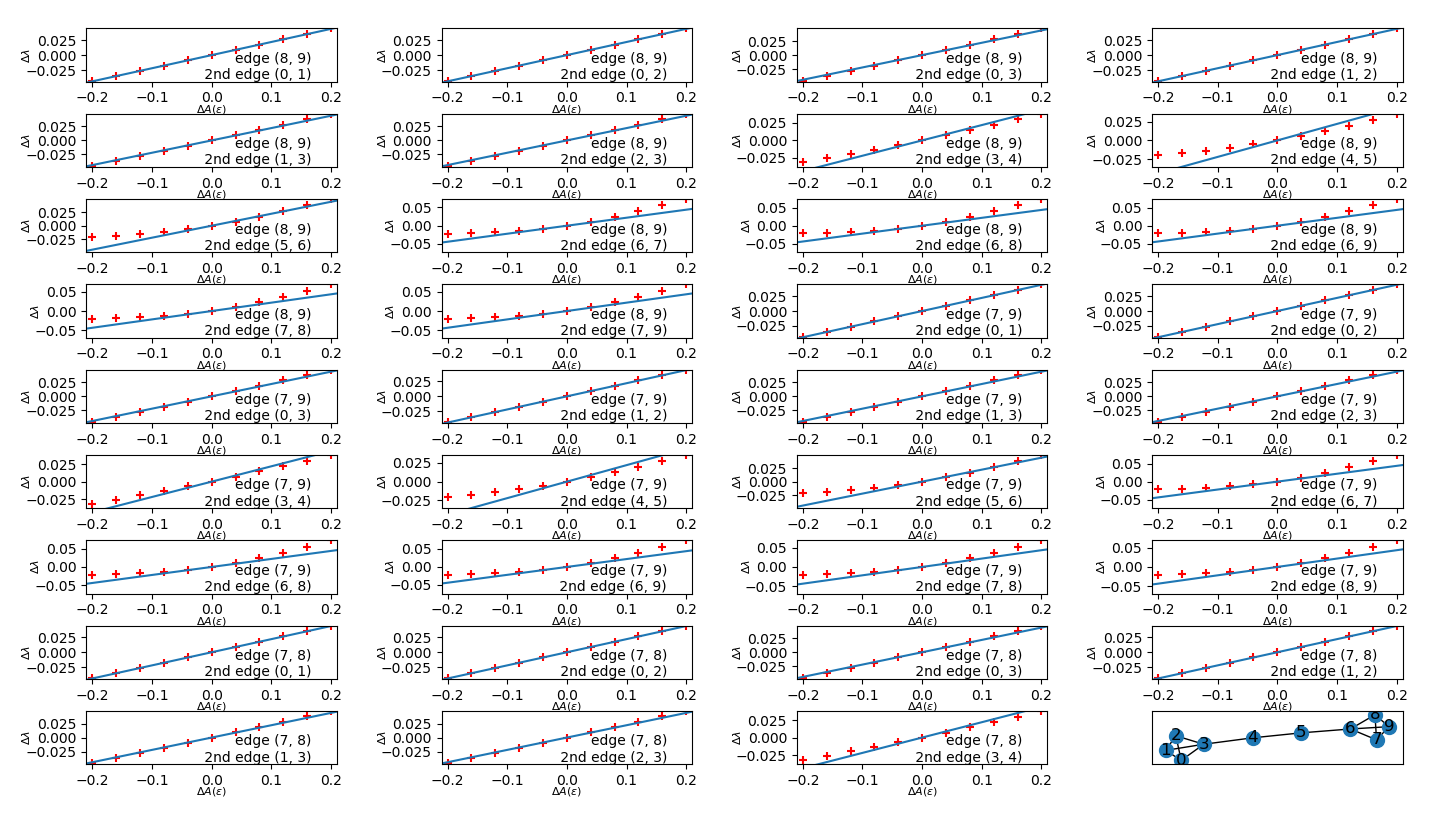}
    \caption{Scatter plot of perturbations $\Delta A$ and the resulting $\Delta \lambda$, compared to line of constant $l_e$, for the case of two edges changing. The plots consider one of the two perturbed edges. Barbell network, with equal initial weights} 
    \label{fig:dual_edge_changes_barbell}
\end{figure}

\begin{figure}[H]
    \centering
    \includegraphics[width = \textwidth]{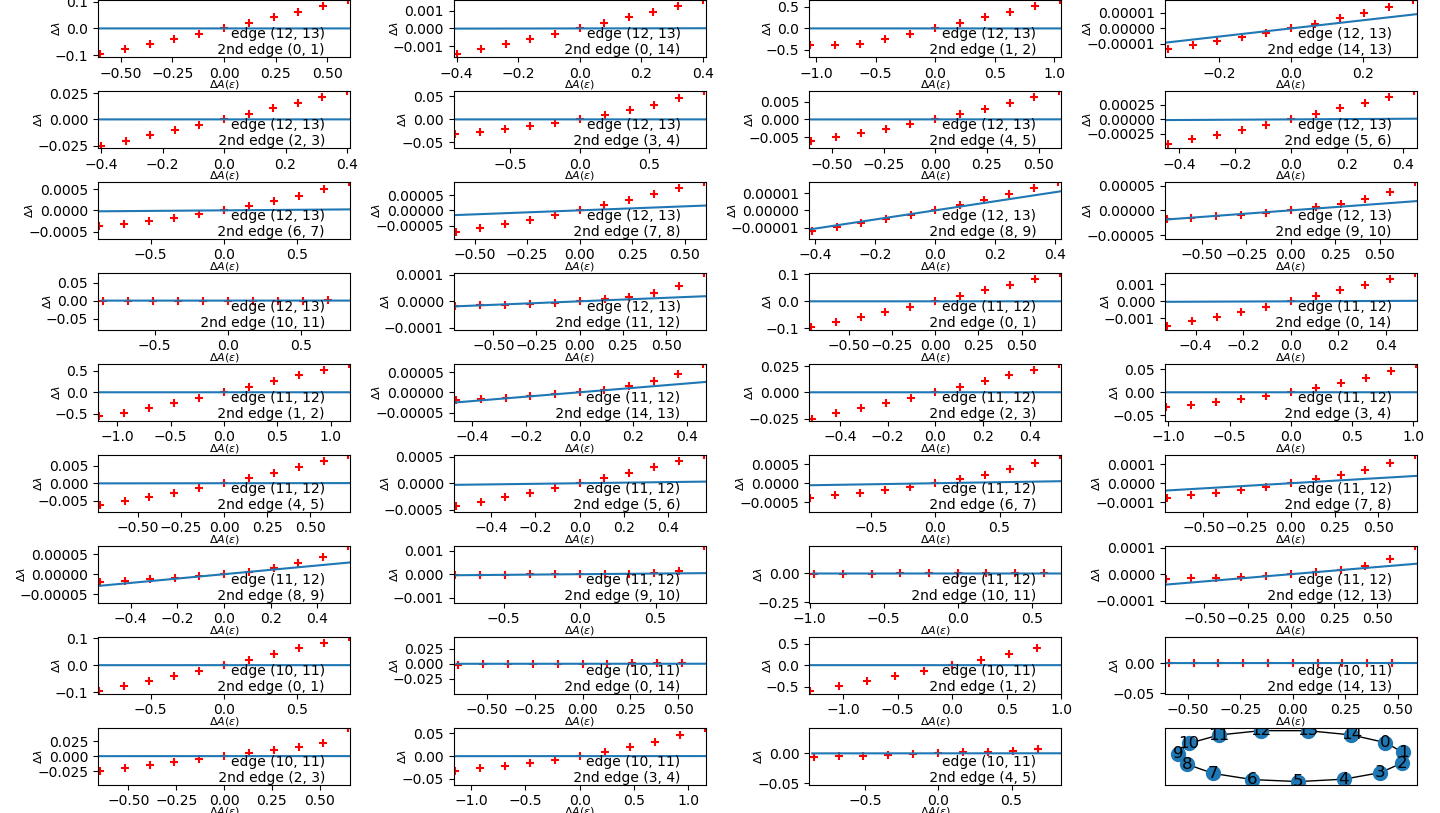}
    \caption{Scatter plot of perturbations $\Delta A$ and the resulting $\Delta \lambda$, compared to line of constant $l_e$, for the case of two edges changing. The plots consider one of the two perturbed edges. Ring network with randomly assigned weights.} 
    \label{fig:dual_edge_changes_ring}
\end{figure}

\begin{figure}[H]
    \centering
    \includegraphics[width = \textwidth]{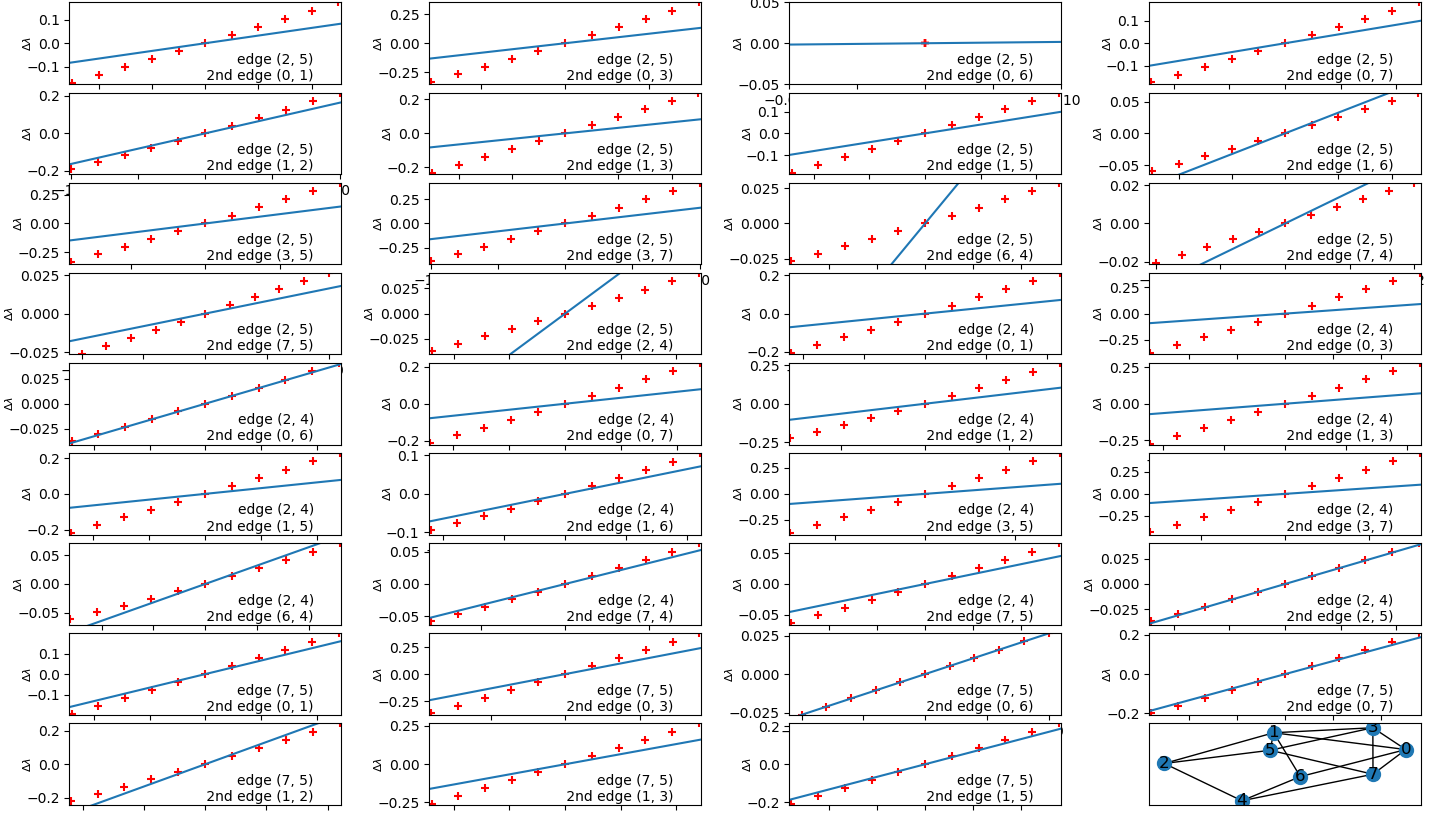}
    \caption{Scatter plot of perturbations $\Delta A$ and the resulting $\Delta \lambda$, compared to line of constant $l_e$, for the case of two edges changing. The plots consider one of the two perturbed edges. Erdős–Rényi network with randomly assigned weights} 
    \label{fig:dual_edge_changes_er}
\end{figure}

\begin{figure}[H]
    \centering
    \includegraphics[width = \textwidth]{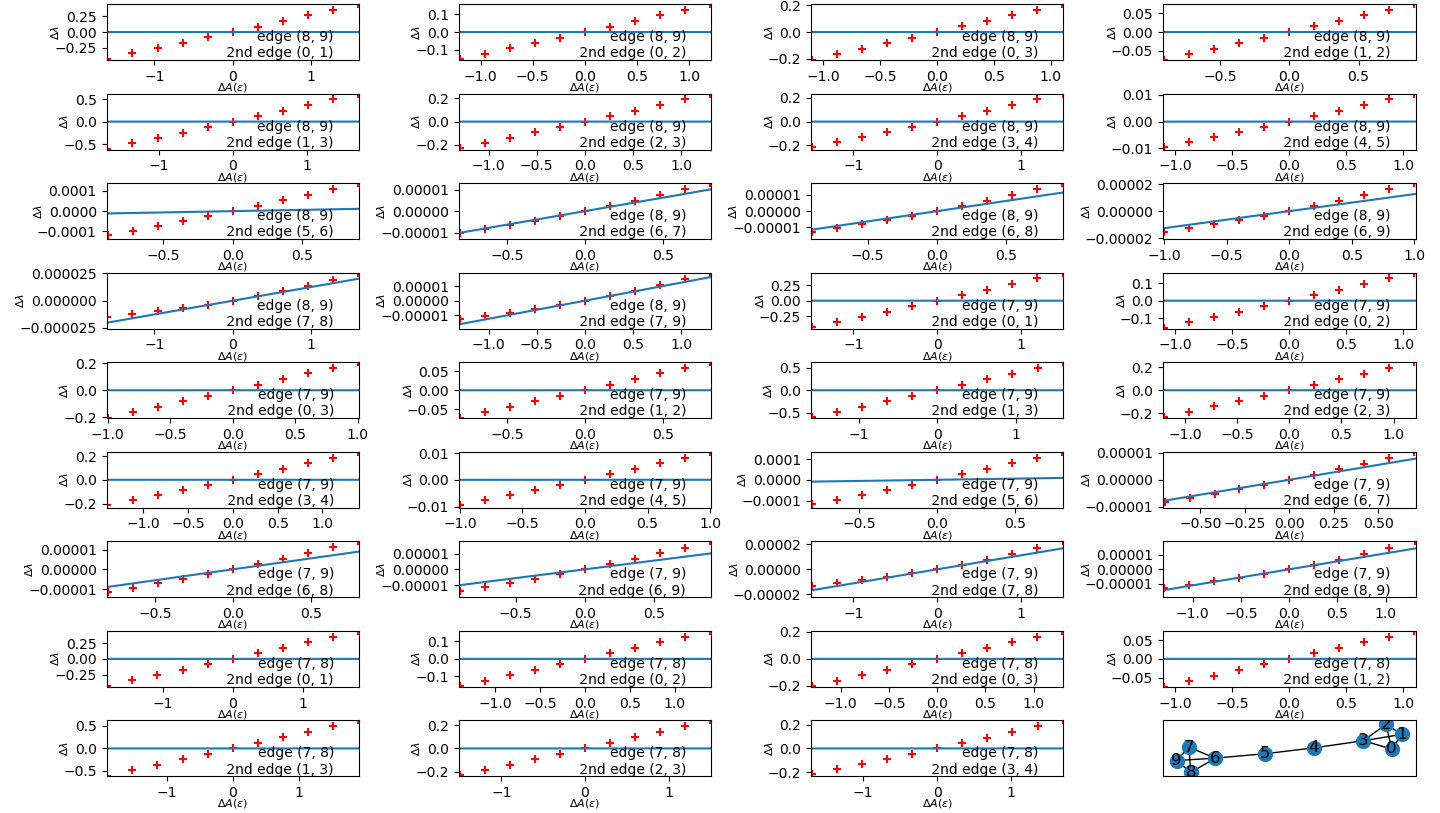}
    \caption{Scatter plot of perturbations $\Delta A$ and the resulting $\Delta \lambda$, compared to line of constant $l_e$, for the case of two edges changing. The plots consider one of the two perturbed edges. Barbell network, with randomly assigned weights.} 
    \label{fig:dual_edge_changes_barbell_w}
\end{figure}

\begin{figure}[H]
    \centering
    \includegraphics[width = \textwidth]{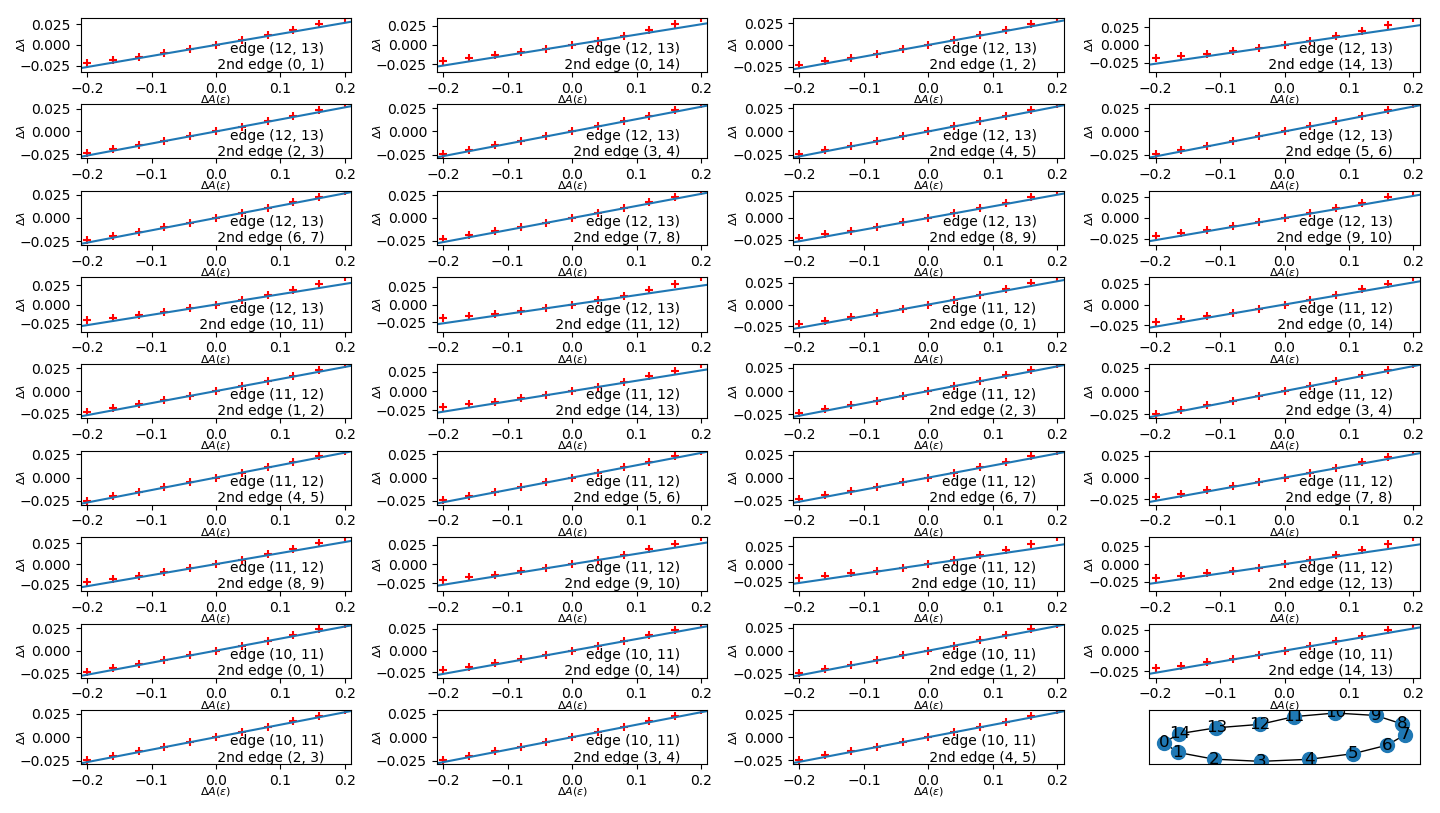}
    \caption{Scatter plot of perturbations $\Delta A$ and the resulting $\Delta \lambda$, compared to line of constant $l_e$, for the case of two edges changing. The plots consider one of the two perturbed edges. Ring network with randomly assigned weights.} 
    \label{fig:dual_edge_changes_ring_w}
\end{figure}

In figures \ref{fig:dual_edge_changes_barbell} - \ref{fig:dual_edge_changes_ring_w} we present the results of changing two individual edges, and observing the resulting change in $\lambda$ for the range of perturbations applied. We overlay this with a line of constant $l_e$, to assess the performance of our approximation. 

\begin{figure}[H]
    \centering
    \includegraphics[width = \textwidth]{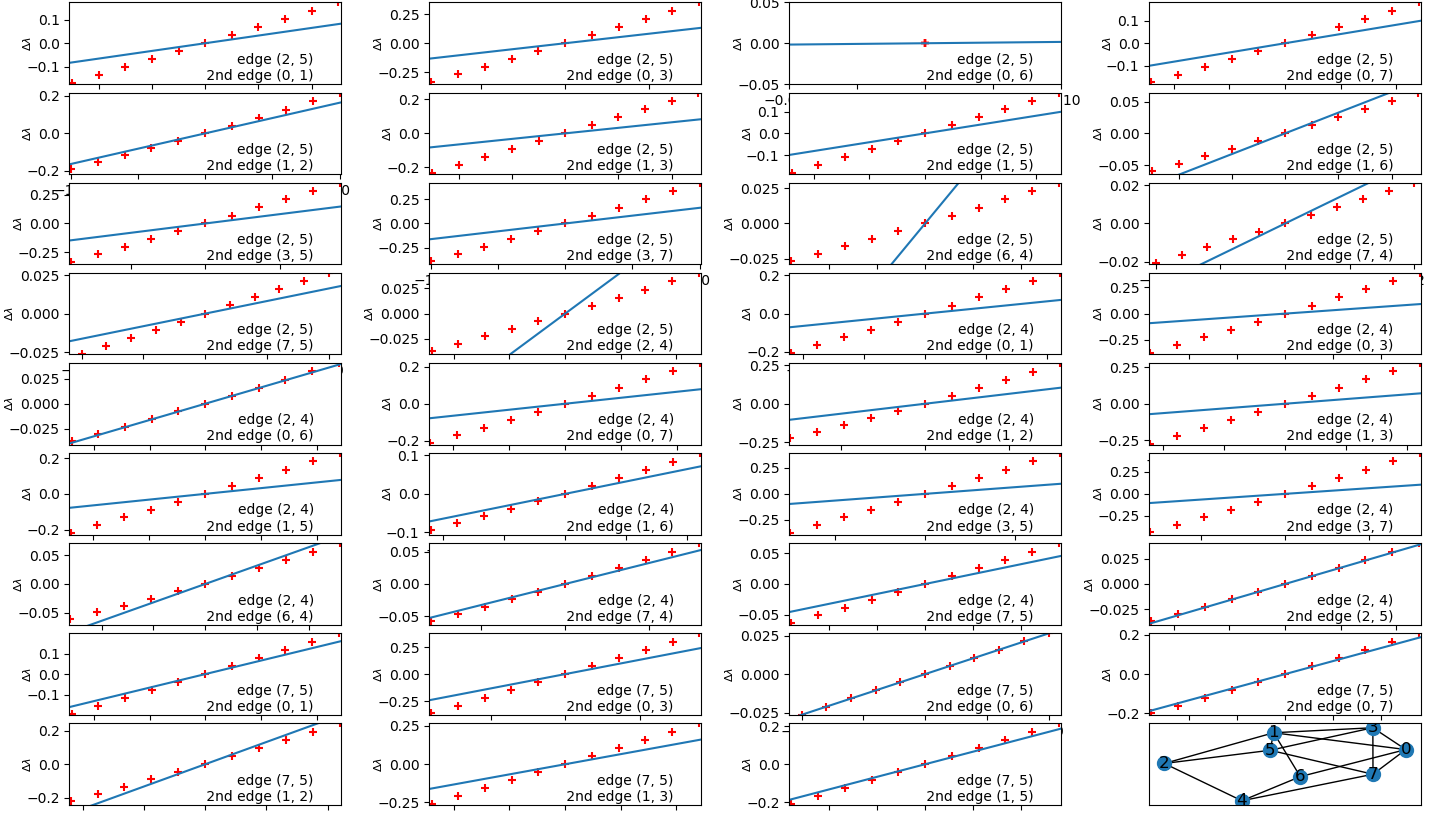}
    \caption{Scatter plot of perturbations $\Delta A$ and the resulting $\Delta \lambda$, compared to line of constant $l_e$, for the case of two edges changing. The plots consider one of the two perturbed edges. Erdős–Rényi network with randomly assigned weights} 
    \label{fig:dual_edge_changes_er_w}
\end{figure}
\section{Perturbation theory approach to deriving network eigenvalue derivatives}
\label{appendix:eig_derivs}

\subsection{Undirected case}
Consider a perturbation to the adjacency matrix $\mathbf{A}$:
\begin{equation}
    \mathbf{A} \rightarrow \mathbf{A}+\epsilon\mathbf{V}
\end{equation}
and the resulting first order changes to the leading eigenvalue $\lambda$ and the associated eigenvector $\ket {\lambda}$:
\begin{equation}
    \lambda = \lambda_0 +\epsilon\lambda
\end{equation}
\begin{equation}
    \ket {\lambda} = \ket {\lambda}_0+\epsilon\ket {\lambda}_1
\end{equation}
Substituting these into our eigenvalue equation
\begin{equation}
    (\mathbf{A}+\epsilon\mathbf{V})(\ket {\lambda}_0+\epsilon\ket {\lambda}_1)=(\lambda_0+\epsilon\lambda_1+...)(\ket {\lambda}_0+\epsilon\ket {\lambda}_1+...)
\end{equation}
and considering terms up to 1st order in $\epsilon$
\begin{equation}
    \mathbf{A}\ket {\lambda}_0+\epsilon\mathbf{V}\ket {\lambda}_0+\epsilon\mathbf{A}\ket {\lambda}_1=\lambda_0\ket {\lambda}_0+\epsilon\lambda_1\ket {\lambda}_0+\epsilon\lambda_0\ket {\lambda}_1
    \label{lin_approx}
\end{equation}
Then we can consider each of the terms in $\epsilon^n$ separately,
\begin{equation}
    \epsilon_0: \textbf{A}\ket {\lambda}_0=\lambda_0\ket {\lambda}_0
\end{equation}
\begin{equation}
    \epsilon_1: \textbf{V}\ket {\lambda}_0+\mathbf{A}\ket {\lambda}_1=\lambda_1 \ket {\lambda}_0+ \lambda_0\ket {\lambda}_1
\end{equation}
By multiplying the equation for $\epsilon^1$ by the left eigenvector $\prescript{}{0}{\bra{\lambda}}$, and making use of the hermitian properties of $\mathbf{A}$ such that $\prescript{}{0}{\bra{\lambda}}\mathbf{A}=\lambda_0\prescript{}{0}{\bra {\lambda}}$, we find
\begin{equation}
    \tensor*[_0]{\braket{\lambda|\mathbf{V}|\lambda}}{_0} = \lambda_1\tensor*[_0]{\braket{\lambda|\lambda}}{_0} 
\end{equation}

The perturbation we are considering is changing one row, and one column, i.e. where $V_{ij}=A_{ij}$ if $i$ or $j$ are the row/column we are changing, zero otherwise:
\[
    V_{ij}=\begin{cases}
    A_{ij} & \textrm{ if } i=k \textrm{ or } j=k\\ 0 & \textrm{ otherwise}
    \end{cases}
\]
Then, expanding the indices, 
\begin{equation}
    \sum_{ij} \lambda_{0,i} V_{ij}\lambda_{0,j} =\sum_{ij}\lambda_{0,i} A_{ij}\lambda_{0,j} \delta_{ik}+ \sum_{ij}\lambda_{0,i} A_{ij}\lambda_{0,j} \delta_{jk} = 2\sum_{j}\lambda_{0,k} A_{kj}\lambda_{0,j}
 \end{equation}
Where we have re-labelled the indices for the second term and have evaluated the $\delta$'s.
Leading us to the result:
\begin{equation}
   l_e = \frac{\partial\lambda}{\partial A_{ij}}=2\lambda_{0,i}\lambda_{0,j}
    \label{undirected l_e}
\end{equation}

where $\lambda_{0,i}$ is the $i$th component of the eigenvector corresponding to the leading eigenvalue of $\mathbf{A}$
\subsection{Directed case}
For the directed case, we still have that 

\begin{equation}
    \epsilon_1: \textbf{V}\ket {\lambda}_1+\mathbf{A}\ket {\lambda}_0=\lambda_1 \ket {\lambda}_0+ \lambda_0\ket {\lambda}_1
\end{equation}
but we can't use the hermitian properties of the matrix $\mathbf{A}$ as for directed networks $\mathbf{A}$ is generally not symmetric. We can however consider the matrix $\mathbf{M} = \mathbf{AA^T}$ and perturbation $\mathbf{M}\rightarrow\mathbf{M}+\epsilon \mathbf{W}$, and use the symmetric result from this. This is useful since the singular values of matrix $\mathbf{A}$ are defined as the square root of the eigenvalues of $\mathbf{AA^T}$, such that:
\begin{equation}
    \partial \lambda^M = 2s^A \partial s^A
\end{equation}
where $\lambda^M$ is the leading eigenvalue of $\mathbf{M}$ and $s^A$ is the leading singular value of $\mathbf{A}$. We can then make use of our result above for the symmetric matrix, 
\begin{equation}
    \tensor*[_0]{\braket{\lambda^M|\mathbf{W}|\lambda^M}}{_0} = \lambda^M_1
\end{equation}
Where $\prescript{}{0}{\bra{\lambda^M}}$ and $\ket {\lambda^M}_0$ are the left and right eigenvectors of $\mathbf{M}$. 
For the directed case, our perturbation is changing just a row (or column) independently, i.e. 
\begin{equation}
    W_{ij}=\begin{cases}
    M_{ij} \textrm{ if } i=k & 0 \textrm{ otherwise}
    \end{cases}
\end{equation}
Then, expanding the indices, 
\begin{equation}
    \sum_{ij} \lambda^M_{0,i} W_{ij}\lambda^M_{0,j} =\sum_{ij}\lambda^M_{0,i} M_{ij}\lambda^M_{0,j} \delta_{ik} = \sum_{j}\lambda^M_{0,k} M_{kj}\lambda^M_{0,j}
 \end{equation}

Leading us to the result

\begin{equation}
    \frac{\partial s^A}{\partial M_{ij}}=\frac{\lambda^M_{0,i} \lambda^M_{0,j}}{2s^A}
    \label{directed l_e}
\end{equation}
where $\lambda^M_{0,i}$ is the $i$th component of the eigenvector corresponding to the leading eigenvalue of $\mathbf{M}$
In both the directed and undirected case above, it is worth noting that the derivations can be generalised to allow new links to be added/removed, however new nodes cannot be added or removed.

\section{Kernel Density Estimation for conditional probability estimation}
\label{appendix:trans_prob}

We make use of multivariate conditional Kernel Density Estimation (KDE) to find the probability distributions for the values of $l_e$. The functional form of the KDE is 
\begin{equation}
    \hat{f_h}(x)=\frac{1}{n}\sum_{i=1}^n K_h(x-x_i) = \frac{1}{nh}\sum_{i=1}^n K\left(\frac{(x-x_i)}{h}\right)
\end{equation}{}
where 
\begin{equation}
    K_h(x)=\frac{1}{h}K\frac{x}{h}
\end{equation}
is the kernel function, a non-negative function. The parameter h is the bandwith, a smoothing parameter. We have used a Gaussian kernel:
\begin{equation}
    K=\frac{1}{\sigma \sqrt{2\pi}}\exp{-\frac{1}{2}\left(\frac{x-\mu}{\sigma}\right)}
\end{equation}

\section{Dataset descriptions}
\label{Appendix_datasets}
The first dataset considered tracks bilateral trade flows between states from 1870-2014, describing import and export data in current U.S. dollars for pairs of sovereign states \cite{COW_data}. This dataset is interesting not just due to its relevance to our focus on financial markets, but also due to an observed growth across time, apart from in two time periods corresponding to the First and Second World Wars.

The second dataset considered was a dataset of private messages sent on an online social network at the University of California. An edge (u, v, t) means that user u sent a private message to user v at time t. As this network is unweighted, the weights of all of the edges have been set to 1. The network was aggregated to daily snapshots, in which the edge weight is the number of times that edge is active during that day. 

Finally, in order to observe the effects of different trading structures on the output of our methods, we applied our techniques to transaction reports relating to three different equity stocks traded on the UK capital markets. The data was aggregated daily, and covers a 2 year period from January 2018\footnote{the Equity-3 dataset was analysed for a 5 month period ending in November 2019}. The Equity-3 dataset was analysed for the shorter time range of 03/06/2019 to 05/11/2019. We chose to study networks of transactions for stocks on energy companies due to the high level of trading activity in is sector. The results displayed in this paper consider the giant component networks of 3 different stocks. The first two instruments were traded without the presence of CCPs, one focusing on oil and gas exploration and production and the second focusing on renewable and alternative energy. The third instrument, another oil and gas production stock, shows a network dominated by the presence of a CCP. Due to the sensitivity of the data, these have been referred to as Equity networks 1, 2 and 3 throughout this paper. 

\subsection{Summary statistics for the Equity datasets}
The Equity data was made available by the FCA for use in this study, and is not publically available. To provide the reader with additional context, here we include some high level network statistics for these networks. All statistics are based on the networks following the removal of nodes which appear on less than 5 days in the sample, which we classed as `inactive'. 

We see that all three networks have similar connectivities, but Equity-3 is significantly denser than the other two and shows a higher level of reciprocity. All datasets show similar values for the correlation coefficient of the adjacency matrix.

\begin{table}[H]
    \centering

    \begin{tabular}{|c|c|c|c|c|c|c|}
         \hline
        dataset & \# nodes & \# edges & connectivity & density & reciprocity & correlation coefficient\\
         \hline
        Equity - 1 & 232 & 6, 961 & 30\% & 12.99\% & 67.72 & 62.9\%\\
        Equity - 2 & 94 & 3,684 & 39.2\% & 42.14\% & 75.03\% & 56.84\% \\
        Equity -3 & 263 & 9.094 & 34.58\% & 13.2\% & 57.66\% & 51.23\% \\
         \hline
    \end{tabular}
    \caption{Network statistics for the three Equity datasets}
    \label{tab:netstats}
\end{table}

\begin{figure*}
    \centering
\begin{subfigure}{.33\textwidth}
     \centering
    \includegraphics[scale=0.33]{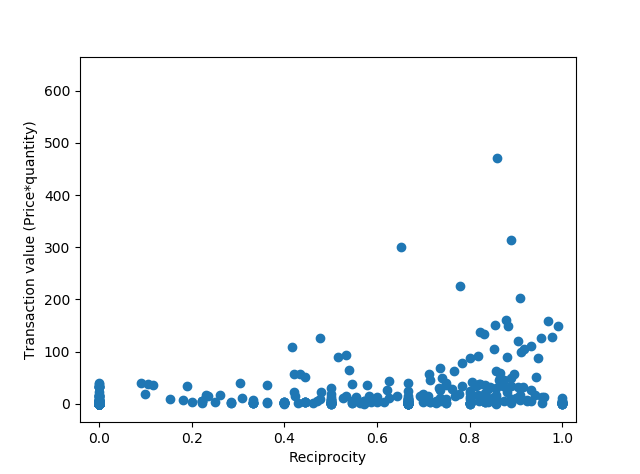}
    \caption{Equity-1}
    \label{fig:reciprocity_green}
\end{subfigure}%
\begin{subfigure}{.33\textwidth}    
   \centering
    \includegraphics[scale=0.33]{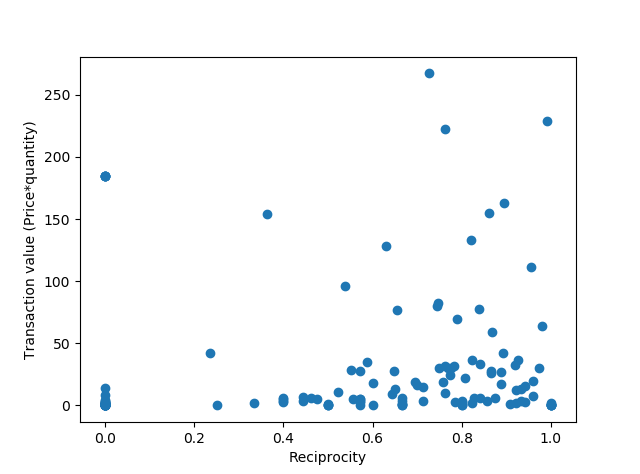}
    \caption{Equity-2}
    \label{fig:reciprocity_brown}
\end{subfigure}
\begin{subfigure}{.33\textwidth}
   \centering
    \includegraphics[scale=0.33]{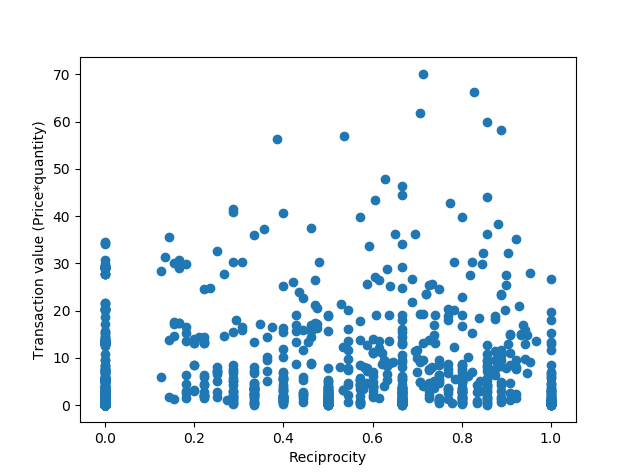}
    \caption{Equity-3}
    \label{fig:reciprocity_vesta}
\end{subfigure}
\caption{Reciprocity vs. price for the Equity networks}

\end{figure*}

We can see from figures \ref{fig:reciprocity_green} and \ref{fig:reciprocity_brown} that large transaction values are more likely to have a high reciprocity for the first and second dataset. However the same cannot be said for the third Equity dataset, as shown in figure \ref{fig:reciprocity_vesta}. 

\begin{figure}
    \centering
    \includegraphics[scale=0.4]{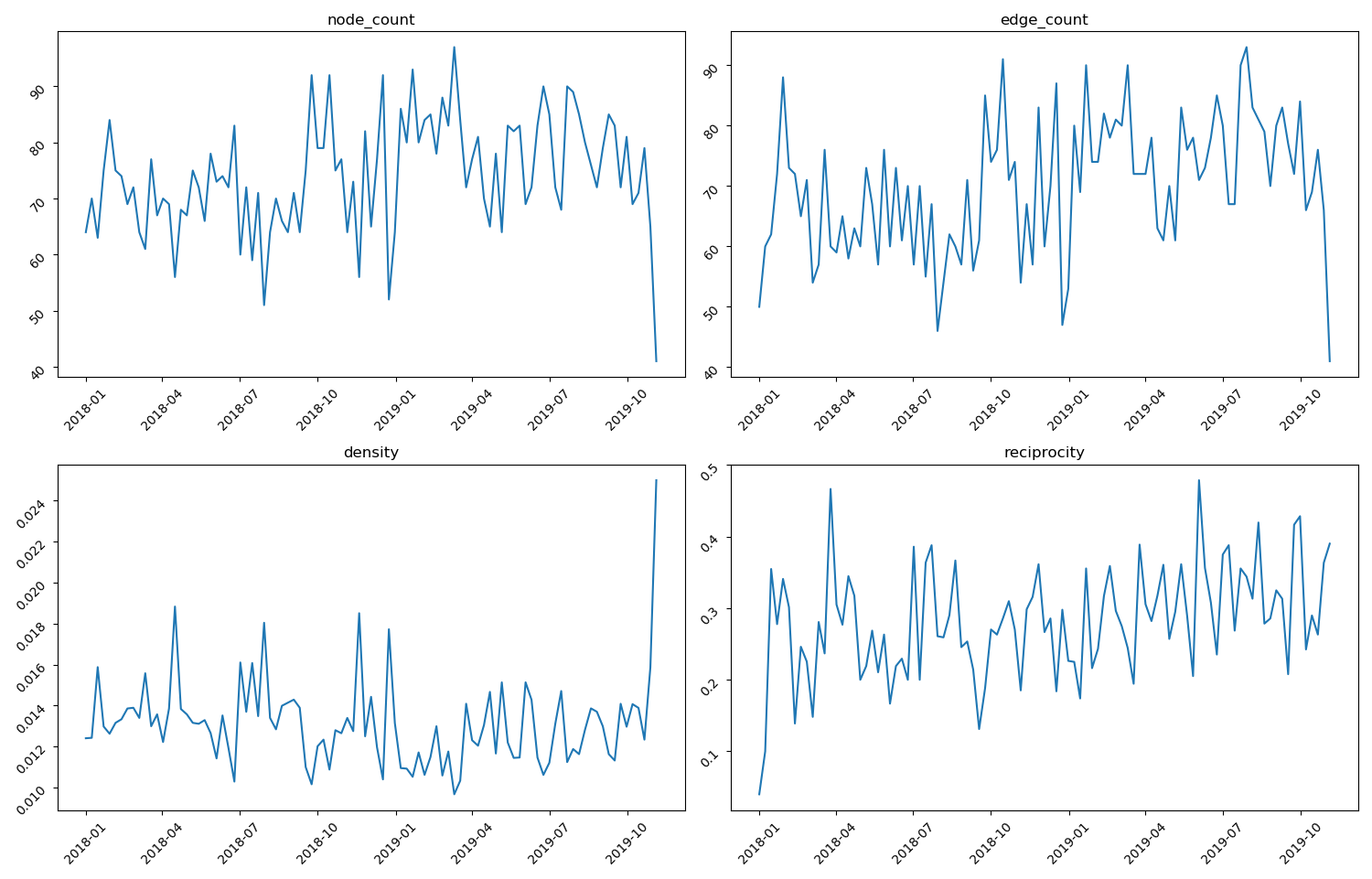}
    \caption{Daily counts of nodes and edges, density and reciprocity across the entire investigation period for Equity-3}
    \label{fig:temp_stats_green}
\end{figure}

\begin{figure}
    \centering
    \includegraphics[scale=0.4]{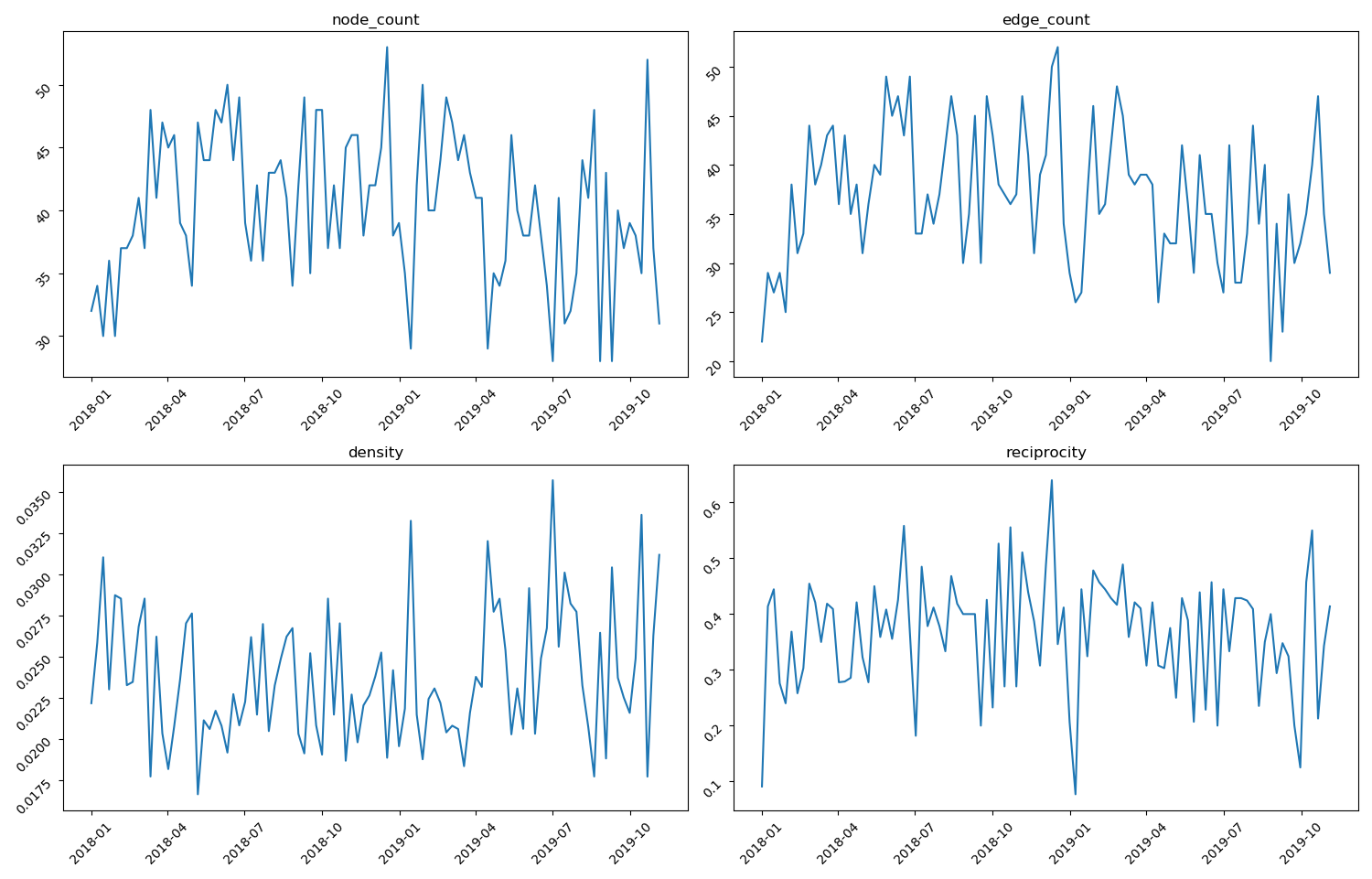}
    \caption{Daily counts of nodes and edges, density and reciprocity across the entire investigation period for Equity-2}
    \label{fig:temp_stats_brown}
\end{figure}

\begin{figure}
    \centering
    \includegraphics[scale=0.4]{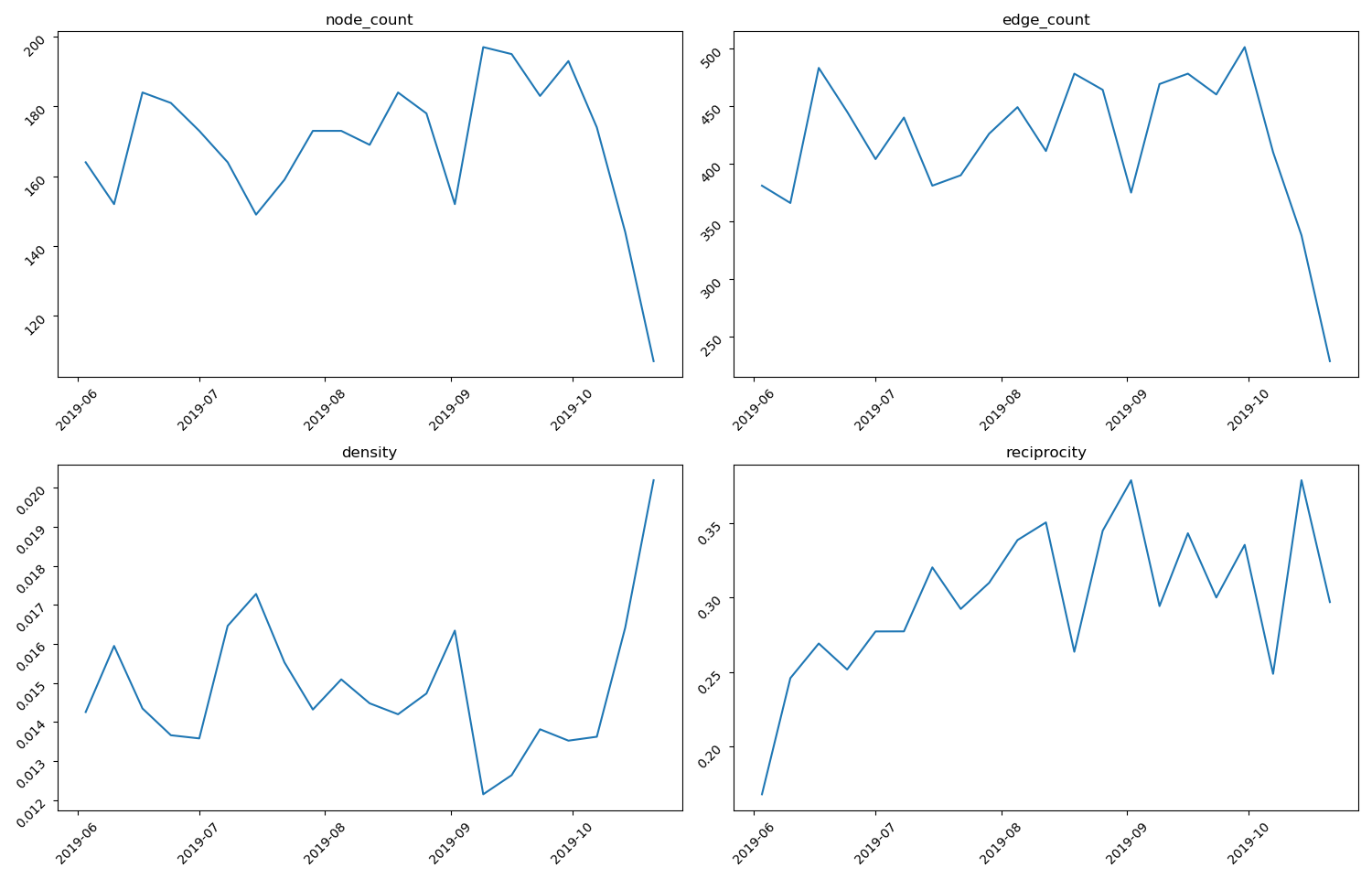}
    \caption{Daily counts of nodes and edges, density and reciprocity across the entire investigation period for Equity-3}
    \label{fig:temp_stats_vestas}
\end{figure}

The evolution of high level network statistics are shown in figures \ref{fig:temp_stats_green},\ref{fig:temp_stats_brown}, and \ref{fig:temp_stats_vestas}. Here we see that the networks fluctuate around a relatively stable mean, with no obvious level of growth or decay across the time period. 


\begin{figure}
    \centering
    \includegraphics[scale=0.35]{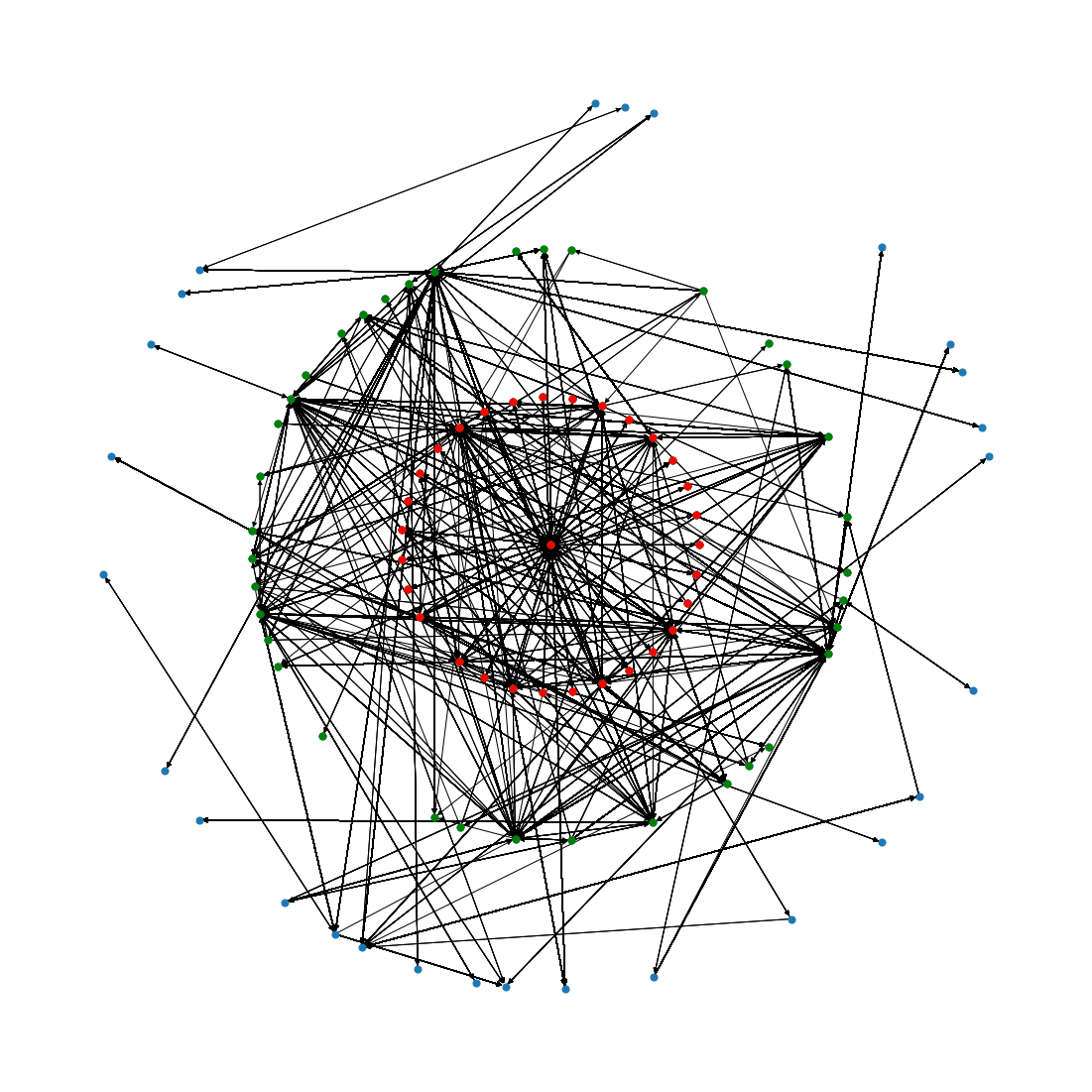}
    \caption{Example of a tiered structure in the GWCC of the trading network for an instrument frequently traded via an individual CCP.}
    \label{fig:ccp}
\end{figure}

It is further interesting to note that the third network considered shows the presence of a  hierarchy in the network, due to it being an instrument that is traded mainly through the use of Central Clearing Parties (CCP), producing a tiered structure as shown in figure \ref{fig:ccp}. Such a structure can be identified following the identification of a dominant node in the network, i.e. a node with significantly higher degree, and examining its ego network. 

\section{Parameter estimations}
\label{parameter_estimations}
\subsection{Estimation of $\rho$ and $\alpha$}
We assume in this section that our networks can be described by a model in which the probability of an edge changing is given by 

\[P(\Delta)=  \theta_e = \left\{
\begin{array}{ll}
      0 & \alpha l_e^{\rho}\leq 0 \\
      \alpha l_e^{\rho} & 0< \alpha l_e^{\rho}< 1 \\
      1 & \alpha l_e^{\rho}\geq 1 \\
\end{array} 
\right. \]
The maximum likelihood estimate of $\theta$ then follows the same procedure as in the case of a (potentially biased) coin toss - given a sample of changes $x_i$, the likelihood of observing these changes given $\theta$ is 
\begin{equation}
    L(x_1, x_2, ...x_n|\theta)=\prod_{e} f(x_e|\theta_e)
\end{equation}
where $f(x_e|\theta_e)$ follows the Bernoulli distribution $\theta_e^{k_e}(1-\theta_e)^{1-k_e}$ where $k_e$ is the observed outcome of edge $e$. Taking the logarithm of this, our log-likelihood is given by 
\begin{equation}
    ln(L(\mathbf{x}|\mathbf{\theta}))=\sum_{e}^N k_e ln(\theta_e) +(1-k_e)ln(1-\theta_e)
\end{equation}
Since $\alpha l_e^{\rho}$ is constrained to be a probability, to estimate the parameters which result in the maximum likelihood, we need to minimise the negative log-likelihood with respect to multiple inequality constraints:
\begin{equation}
    0\leq \alpha l_e^{\rho} \leq 1
\end{equation}
Where we have one inequality constraint for each $l_e$. To do this, we make use of the Karush-Kuhn-Tucker conditions \cite{KKT} and numerical optimisation, to find the optimal saddle point which maximises $L$ with whilst satisfying these constraints.  

\subsection{Estimation of $\beta$ and $\gamma$}
For the case of the distribution of edge changes drawn from a Gaussian distribution with $\mu$=0 and $\sigma=\beta l_e^{\gamma~}$, the log-likelihood is given by
\begin{equation}
    \ln(L) =\sum_e^N \ln\left(\frac{1}{\sqrt{2\pi} \beta l_e^{\gamma}}\right) \exp{\left(\frac{-x_e^2}{2\beta^2 l_e^{2\gamma}}\right)}
\end{equation}
where $x_e$ refers to the observed relative change of edge $e$.
Differentiating with respect to $\beta$, 
\begin{equation}
    \beta=\sqrt{\frac{1}{N}\sum_e^N \frac{x_e^2}{l_e^{2\gamma}}}
    \label{beta_appdx}
\end{equation}
From which we recover the expected standard deviation for a Gaussian in the case of $\gamma$=0. 

Differentiating with respect to $\gamma$, 
\begin{equation}
    \frac{\partial\ln(L)}{\partial\gamma} = \sum_e^N -\ln(l_e)+ \frac{\partial}{\partial \gamma}\frac{x_e^2}{2\beta^2} \exp{(-2\gamma \ln{l_e})}
\end{equation}
which when set to 0,
\begin{equation}
    \sum_e^N \ln(l_e)\left(1+ \frac{x_e^2 ln(l_e)}{\beta^2 l_e^{2\gamma}}\right)
\end{equation}
Substituting \ref{beta_appdx} for $\beta$, and solving numerically allows us to produce an estimate for $\gamma$, which can be solved numerically for $x_e=\Delta A_{rel}$. 

\section{Comparison of data distributions to model}
\begin{figure}[H]
    \centering
    \includegraphics[width=\textwidth]{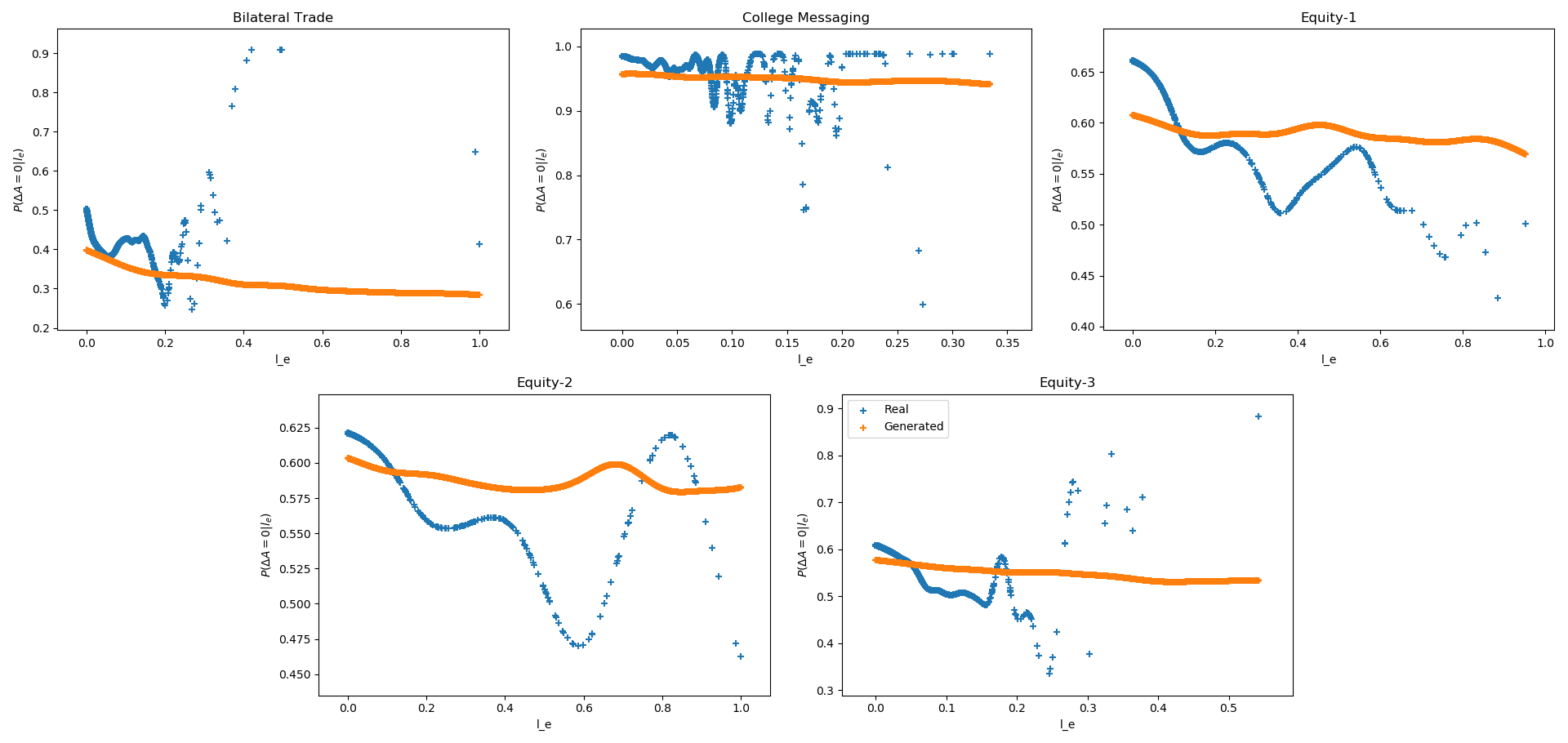}     \caption{$P(\Delta A=0|\ln(l_e))$ as a function of $\ln(l_e)$ for the 5 real datasets, overlaid with the distributions for data generated according to the model in \ref{markov_chain}}
    \label{fig:param_all}
\end{figure}

Figure \ref{fig:param_all} shows the bulk of the distributions for $P(\Delta A=0|\ln(l_e))$ for our 5 datasets, in comparison to the equivalent generated from our model for network evolution given by equation \ref{markov_chain}.
\end{document}